\documentclass{aa}
\usepackage{graphicx}

\def\mb#1{\setbox0=\hbox{$#1$}\kern-.025em\copy0\kern-\wd0
\kern-0.05em\copy0\kern-\wd0\kern-.025em\raise.0233em\box0}

\begin{document}
   \title{On the lifetime of metastable states in \\
self-gravitating systems}

   \author{P.H. Chavanis}

\institute{ Laboratoire de Physique Th\'eorique, Universit\'e Paul
Sabatier, 118
route de Narbonne 31062 Toulouse, France\\
\email{chavanis@irsamc.ups-tlse.fr}}

   \date{To be included later }

   \abstract{We discuss the physical basis of the statistical
   mechanics of self-gravitating systems.  We show the correspondance
   between statistical mechanics methods based on the evaluation of
   the density of states and partition function and thermodynamical
   methods based on the maximization of a thermodynamical potential
   (entropy or free energy). We address the question of the
   thermodynamic limit of self-gravitating systems, the justification
   of the mean-field approximation, the validity of the saddle point
   approximation near the transition point, the lifetime of metastable
   states and the fluctuations in isothermal spheres. In particular,
   we emphasize the tremendously long lifetime of metastable states of
   self-gravitating systems which increases exponentially with the
   number of particles $N$ except in the vicinity of the critical
   point. More specifically, using an adaptation of the Kramers
   formula justified by a kinetic theory, we show that the lifetime of
   a metastable state scales as $e^{N\Delta s}$ in microcanonical
   ensemble and $e^{N\Delta j}$ in canonical ensemble, where $\Delta
   s$ and $\Delta j$ are the barriers of entropy and free energy
   $j=s-\beta \epsilon$ (per particle) respectively. The physical
   caloric curve must take these metastable states (local entropy
   maxima) into account. As a result, it becomes multi-valued and
   leads to microcanonical phase transitions and ``dinosaur's necks''
   (Chavanis 2002b, Chavanis \& Rieutord 2003). The consideration of
   metastable states answers the critics raised by D.H.E. Gross
   [cond-mat/0307535/0403582].

   \keywords{stellar systems: theory; statistical mechanics.
               }
   }

   \maketitle
%

\section{Introduction}
\label{sec_introduction}

The statistical mechanics of self-gravitating systems has a long
history starting with the seminal papers of Antonov (1962) and
Lynden-Bell \& Wood (1968). A statistical mechanics approach is
particularly relevant to describe the late stages of ``small'' groups
of stars ($N\sim 10^{6}$), such as globular clusters, which evolve
under the influence of stellar encounters (``collisional''
relaxation). Apart from astrophysical applications, the statistical
mechanics of stellar systems is of great interest in physics because
it differs in many respects from that of more familiar systems with
short-range interactions (Padmanabhan 1990).  In particular, for
systems with long-range interactions, the thermodynamical ensembles
are not equivalent, negative specific heats are allowed in the
microcanonical ensemble (but not in the canonical ensemble) and
metastable equilibrium states can have tremendously long lifetimes
making them of considerable interest.

Two types of approaches have been developed to determine the
statistical equilibrium state of a self-gravitating system. In the
{\it thermodynamical approach}, one determines the { most probable}
distribution of particles by maximizing the Boltzmann entropy at fixed
mass and energy in the microcanonical ensemble or by minimizing the
free energy $F=E-TS$ at fixed mass and temperature in the canonical
ensemble (Lynden-Bell \& Wood 1968, Katz 1978, Chavanis 2002a). This
approach is the simplest and the most illuminating. In addition, it is
directly related to kinetic theories (based on the Landau or on the
Fokker-Planck equation) for which the Boltzmann entropy (or the
Boltzmann free energy) plays the role of a Lyapunov functional and
satisfies a H-theorem. Alternatively, in the {\it statistical
mechanics approach}, one starts from the density of states or
partition function, transforms it into a functional integral and uses
a saddle point approximation valid in a properly defined thermodynamic
limit (Horwitz
\& Katz 1978, de Vega \& Sanchez 2002, Katz 2003).

In the first part of this paper, we discuss the connexion between
these two procedures. We remain at a heuristic level, stressing more
the physical ideas than the mathematical formalism. In
Sec. \ref{sec_fermions}, we introduce the entropy by a combinatorial
analysis. In order to regularize the problem at short distances, we
consider either the case of self-gravitating fermions or the case of
self-gravitating particles with a soften potential. We also discuss
the thermodynamic limit of the classical and quantum self-gravitating
gas. In Sec. \ref{sec_connexion}, we show the relation between the
density of states $g(E)$ and the entropy functional $S[f]$ and between
the partition function $Z(\beta)$ and the free energy functional
$J[f]=S[f]-\beta E[f]$. In the thermodynamic limit, the saddle point
approximation amounts to maximizing the entropy at fixed mass and
energy (microcanonical ensemble) or to minimizing the free energy at
fixed mass and temperature (canonical ensemble). In
Sec. \ref{sec_corr}, we discuss the notion of canonical and
microcanonical phase transitions in self-gravitating systems.  We
perform the (standard) horizontal and (less standard) vertical Maxwell
constructions and discuss the validity of the saddle point
approximation near the transition point for finite $N$ systems. These
results (e.g., microcanonical first order phase transitions) are
relatively new in statistical mechanics and still subject to
controversy (Gross 2003,2004). Therefore, we provide a relatively
detailed discussion of these issues.

In the second part of the paper, we emphasize the importance of
metastable states in astrophysics and show how they can be taken
into account in the statistical approach. In Sec.
\ref{sec_persistence}, we use the Kramers formula to estimate the
lifetime of a metastable state. We show that the lifetime of a
metastable state scales as $e^{N\Delta s}$ in microcanonical
ensemble and $e^{N\Delta j}$ in canonical ensemble, where $\Delta
s$ and $\Delta j$ are the barriers of entropy and free energy
$j=s-\beta \epsilon$ (per particle) respectively. Therefore, the
typical lifetime of a metastable state scales as $e^{N}$ except in
the vicinity of the critical point $E_{c}$ (Antonov energy) or
$T_{c}$ (Emden-Jeans temperature). We explicitly compute the barriers of
entropy and free energy close to the critical point for classical
self-gravitating particles (stars). The very long lifetime of
metastable states, scaling as $e^{N}$, was pointed out by Chavanis
\& Rieutord (2003) and the difficulty of a stellar system to overcome
the entropic barrier and collapse was qualitatively discussed in
Chavanis \& Sommeria (1998). We here improve these arguments by
developing a theory of fluctuations in isothermal spheres, following
the approach of Katz \& Okamoto (2000). We also determine how
finite $N$ effects affect the collapse temperature and the 
collapse energy. Finally, in Sec. \ref{sec_kramers}, we derive
a Fokker-Planck equation for the evolution of the distribution of
energies $P(E,t)$ in the canonical ensemble and make contact with
the standard Kramers problem. We determine the typical lifetime of
a metastable state by calculating the escape time accross a
 barrier of free energy.

\section{The most probable distribution}
\label{sec_fermions}

\subsection{The Fermi-Dirac distribution}
\label{sec_fd}

We consider a system of $N$ particles confined within a spherical box
of radius $R$ and interacting via Newtonian gravity. Let $f({\bf
r},{\bf v},t)$ denote the distribution function of the system,
i.e. $f({\bf r},{\bf v},t)d^{3}{\bf r} d^{3}{\bf v}$ gives the mass of
particles whose position and velocity are in the cell $({\bf r},{\bf
v};{\bf r}+d^{3}{\bf r},{\bf v}+d^{3}{\bf v})$ at time $t$. The
integral of $f$ over the velocity determines the spatial density
\begin{equation}
\rho=\int f \ d^{3}{\bf v}, \label{fd1}
\end{equation}
and the total mass of the configuration is given by
\begin{equation}
M=\int \rho \ d^{3}{\bf r}, \label{fd2}
\end{equation}
where the integral extends over the entire domain. On the other hand,
in the meanfield approximation, the total energy of the system can be
expressed as
\begin{equation}
E={1\over 2}\int fv^{2}d^{3}{\bf r}d^{3}{\bf v}+{1\over
2}\int\rho\Phi d^{3}{\bf r}=K+W, \label{fd3}
\end{equation}
where $K$ is the kinetic energy and $W$ the potential energy. The
meanfield expression of the potential energy is obtained from the
exact expression
\begin{eqnarray}
W=\biggl \langle -{1\over 2}\sum_{i\neq j}{Gm^{2}\over |{\bf r}_{i}-{\bf r}_{j}|}\biggr \rangle\qquad\qquad\qquad\qquad\nonumber\\
=-{1\over 2}GN(N-1)m^{2}\int {P_{2}({\bf r}_{1},{\bf r}_{2})\over |{\bf r}_{1}-{\bf r}_{2}|}d^{3}{\bf r}_{1}d^{3}{\bf r}_{2}, \label{fd3ex}
\end{eqnarray}
by approximating the two-body distribution function $P_{2}({\bf
r}_{1},{\bf r}_{2})$ by the product of two one-body distribution
functions $P_{1}({\bf r}_{1}){\times} P_{1}({\bf r}_{2})$ and using
$\rho({\bf r})=Nm P_{1}$. For self-gravitating systems, this
mean-field approximation is exact in a proper thermodynamic limit
$N\rightarrow +\infty$ with $\eta=\beta GMm/R$ and
$\Lambda=-ER/GM^{2}$ fixed (see Appendix \ref{sec_bbgky}).  The
gravitational potential $\Phi=-G\int
\rho({\bf r}')/|{\bf r}-{\bf r}'|d^{3}{\bf r}'$ is solution of the
Newton-Poisson equation
\begin{equation}
\Delta\Phi=4\pi G\rho. \label{fd4}
\end{equation}

In order to regularize the problem at short distances, we shall invoke
quantum mechanics and use the Pauli exclusion principle. The Pauli
exclusion principle is a fundamental concept in physics and it has
also applications in astrophysics, e.g. in white dwarf and neutron
stars. Therefore, it can be considered as a physically relevant
small-scale regularization for compact objects. We wish to determine
the most probable distribution of self-gravitating fermions at
statistical equilibrium. To that purpose, we divide the individual
phase space $\lbrace {\bf r},{\bf v}\rbrace$ into a very large number
of microcells with size $(h/m)^3$ where $h$ is the Planck constant
(the mass $m$ of the particles arises because we use ${\bf v}$ instead
of ${\bf p}$ as a phase space coordinate). A microcell is occupied
either by $0$ or $1$ fermion (or $g=2s+1$ fermions if we account for
the spin). We shall now group these microcells into macrocells each of
which contains many microcells but remains nevertheless small compared
to the phase-space extension of the whole system. We call $\nu$ the
number of microcells in a macrocell. Consider the configuration
$\lbrace n_i \rbrace$ where there are $n_1$ fermions in the $1^{\rm
st}$ macrocell, $n_2$ in the $2^{\rm nd}$ macrocell etc., each
occupying one of the $\nu$ microcells with no cohabitation. The number
of ways of assigning a microcell to the first element of a macrocell
is $\nu$, to the second $\nu -1$ etc.  Since the particles are
indistinguishable, the number of ways of assigning microcells to all
$n_i$ particles in a macrocell is thus
\begin{equation}
{1\over n_i!}{\times} {\nu!\over (\nu-n_i)!}. \label{fd5}
\end{equation}
To obtain the number of microstates corresponding to the macrostate
$\lbrace n_i \rbrace$ defined by the number of fermions $n_i$ in each
macrocell (irrespective of their precise position in the cell), we
need to take the product of terms such as (\ref{fd5}) over all
macrocells. Thus, the number of microstates corresponding to the
macrostate $\lbrace n_i \rbrace$, i.e. the probability of the state
$\lbrace n_i \rbrace$, is
\begin{equation}
W(\lbrace n_i \rbrace)=\prod_i {\nu!\over n_i!(\nu-n_i)!}.
\label{fd6}
\end{equation}
This is the Fermi-Dirac statistics. As is customary, we define the
entropy of the state $\lbrace n_i \rbrace$ by
\begin{equation}
S(\lbrace n_i \rbrace)=\ln W(\lbrace n_i \rbrace). \label{fd7}
\end{equation}
It is convenient here to return to a representation in terms of
the distribution function giving the phase-space density in the
$i$-th macrocell
\begin{equation}
f_i=f({\bf r}_i,{\bf v}_i)={n_i \ m\over \nu \ ({h\over
m})^3}={n_i\eta_0\over \nu}, \label{fd8}
\end{equation}
where we have defined $\eta_0=m^{4}/h^3$, which represents the
maximum value of $f$ due to Pauli's exclusion principle. Now,
using the Stirling formula, we have
\begin{eqnarray}
\ln W(\lbrace n_i \rbrace)\simeq \sum_i \nu
(\ln\nu-1)-\nu\biggl\lbrace {f_i\over \eta_0}\biggl\lbrack
\ln\biggl ({\nu f_i\over \eta_0}\biggr )-1\biggr\rbrack \nonumber\\
+\biggl
(1-{f_i\over \eta_0}\biggr )\biggl\lbrack\ln\biggl\lbrace
\nu\biggl (1-{f_i\over \eta_0}\biggr
)\biggr\rbrace-1\biggr\rbrack\biggr\rbrace. \label{fd9}
\end{eqnarray}
Passing to the continuum limit $\nu\rightarrow 0$, we obtain the
usual expression of the Fermi-Dirac entropy
\begin{eqnarray}
S=-k_B\int \biggl\lbrace {f\over\eta_{0}}\ln
{f\over\eta_{0}}+\biggl (1-  {f\over\eta_{0}}\biggr)\ln \biggl (1-
{f\over\eta_{0}}\biggr)\biggr\rbrace\ {d^{3}{\bf r}d^{3}{\bf
v}\over ({h\over m})^3}.\nonumber\\
\label{fd10}
\end{eqnarray}
If we take into account the spin of the particles, the above
expression remains valid but the maximum value of the distribution
function is now  $\eta_{0}=g m^{4}/h^{3}$, where $g=2s+1$ is the
spin multiplicity of the quantum states (the phase space element
has also to be multiplied by $g$). In the non-degenerate (or
classical) limit $f\ll\eta_0$, the Fermi-Dirac entropy
(\ref{fd10}) reduces to the Boltzmann entropy
\begin{equation}
S=-k_B\int {f\over m}\biggl\lbrack \ln\biggl ({f h^3\over g
m^{4}}\biggr )-1\biggr \rbrack d^{3}{\bf r}d^{3}{\bf v}.
\label{fd11}
\end{equation}

Now that the entropy has been precisely justified, the statistical
equilibrium state (most probable state) of self-gravitating
fermions is obtained by maximizing the Fermi-Dirac entropy (\ref{fd10}) at
fixed mass (\ref{fd2}) and energy (\ref{fd3}):
\begin{equation}
{\rm Max}\quad S[f]\quad | \quad E[f]=E,\  M[f]=M. \label{fd12}
\end{equation}
Introducing Lagrange multipliers
$1/T$ (inverse temperature) and $\mu$ (chemical potential)
to satisfy these constraints, and writing the variational
principle in the form
\begin{equation}
\delta S-{1\over T}\ \delta E+{\mu\over T} \delta N=0, \label{fd13}
\end{equation}
we find that the {\it critical points} of entropy
correspond to the Fermi-Dirac distribution
\begin{equation}
f={\eta_{0}\over 1+\lambda e^{\beta m ({v^{2}\over 2}+\Phi)}},
\label{fd14}
\end{equation}
where $\lambda=e^{-\beta \mu}$ is a strictly positive constant
(inverse fugacity) and $\beta={1\over k_B T}$ is the inverse
temperature. Clearly, the
distribution function satisfies $f\le \eta_{0}$, which is a
consequence of Pauli's exclusion principle.

So far, we have assumed that the system is isolated so that the energy
is conserved. If now the system is in contact with a thermal bath
(e.g., a radiation background) fixing the temperature, the statistical
equilibrium state minimizes the free energy $F=E-TS$, or maximizes the
Massieu function $J=S-\beta E$, at fixed mass and temperature:
\begin{equation}
{\rm Max}\quad J[f]\quad |\quad M[f]=M. \label{fd15}
\end{equation}
Introducing Lagrange multipliers and
writing the variational principle in the form
\begin{equation}
\delta J+{\mu\over T} \delta N=0, \label{fd16}
\end{equation}
we find that the {\it critical points} of free energy are again given
by the Fermi-Dirac distribution (\ref{fd14}). Therefore, the critical
points (first variations) of the variational problems (\ref{fd12}) and
(\ref{fd15}) are the same. However, the stability of the system
(regarding the second variations) can be different in microcanonical
and canonical ensembles (see, e.g., Chavanis 2002b). When this happens, we
speak of a situation of {\it ensemble inequivalence}. The stability of
the system can be determined by a graphical construction, by simply
plotting the series of equilibria $\beta(E)$ and using the turning
point method of Katz (1978, 2003). Inequivalence of statistical
ensembles occurs when the series of equilibria presents turning points
or bifurcations.

\subsection{Classical particles with soften gravitational potential}
\label{sec_soften}

We now consider a system of classical self-gravitating particles, like
stars in globular clusters. In order to make the problem of
statistical mechanics well-posed mathematically (see below), we
introduce a soften potential of the form
\begin{equation}
u({\bf r}-{\bf r}')={-Gm^{2}\over \sqrt{({\bf r}-{\bf r}')^{2}+r_{0}^{2}}},
\label{a1}
\end{equation}
where $r_{0}$ is the soften radius. As we shall see, the soften radius
$r_{0}$ plays a role similar to the inverse of $\eta_{0}$, the maximum
phase space density, in the case of self-gravitating fermions. As said
previously, this soften radius is introduced in order to pose the
problem correctly. However, we shall argue in the sequel that this
small-scale cut-off is irrelevant for the structure of stellar
systems.

We wish to determine the most probable distribution of stars at
statistical equilibrium (Ogorodnikov 1965). To that purpose, we divide
the individual phase space $\lbrace {\bf r},{\bf v}\rbrace$ into a
very large number of microcells with size $(h/m)^{3}$ where $h$ is a
constant with dimension of angular momentum. Of course, quantum
mechanics is not relevant for stellar systems so that $h$ should not
be confused with the Planck constant in the present context. For
classical systems, a microcell can be occupied by an arbitrary number
of particles. Adapting the counting analysis of Sec.  \ref{sec_fd} to
the present context, the number of microstates corresponding to the
macrostate $\lbrace n_i \rbrace$, i.e. its probability, is
\begin{equation}
W(\lbrace n_i \rbrace)=N!\prod_i {\nu^{n_{i}}\over n_i!}.
\label{a2}
\end{equation}
This is the Maxwell-Boltzmann statistics. If we define the
entropy of the state $\lbrace n_i \rbrace$ by
\begin{equation}
S(\lbrace n_i \rbrace)=\ln W(\lbrace n_i \rbrace), \label{a3}
\end{equation}
and take the continuum limit, we obtain the
usual expression of the Boltzmann entropy
\begin{equation}
S=-k_{B} \int {f\over m}\ln\biggl ({f h^3\over N
m^{4}}\biggr ) d^{3}{\bf r}d^{3}{\bf v}.
\label{a4}
\end{equation}
Note that it differs from the expression (\ref{fd11}) obtained from
the Fermi-Dirac entropy. This is of course related to the Gibbs
paradox in standard thermodynamics (Huang 1963). In the absence of
self-gravity, Eq. (\ref{a4}) reduces to the awkward expression
\begin{equation}
S=Nk_{B}\ln\biggl \lbrack V\biggl ({4\pi m\over 3h^{2}}{E\over N}\biggr )^{3/2}\biggr\rbrack+{3\over 2}N,
\label{a5}
\end{equation}
which is clearly non-extensive. By constrast, Eq. (\ref{fd11}) leads to the
Sackur-Tetrode formula
\begin{equation}
S=Nk_{B} \ln\biggl \lbrack {V\over N}\biggl ({4\pi m\over 3h^{2}}{E\over N}\biggr )^{3/2}\biggr\rbrack+{5\over 2}N,
\label{a6}
\end{equation}
which is extensive. As is well-known, the origin of this discrepency
is due to the indiscernability of the particles and to the presence of
the factor $N!$ in the Maxwell statistics (\ref{a2}). For a molecular
gas, the Gibbs paradox is usualy solved by invoking quantum
mechanics. For a system of stars, one cannot use this argument. We
shall consider that the stars are discernable and use the
expression (\ref{a4}) for the entropy. However, this choice does not
affect the {\it structure} of the equilibrium state as we shall see in
the sequel.

The most probable distribution of stars at statistical equilibrium is
now obtained by maximizing the Boltzmann entropy (\ref{a4}) at fixed
mass and energy. This yields the Maxwell-Boltzmann distribution
\begin{equation}
f=A e^{\beta m ({v^{2}\over 2}+\Phi)}, \label{a7}
\end{equation}
where $\Phi$ is related to the density $\rho$ by
\begin{equation}
\Phi=-G\int {\rho({\bf r}')\over \sqrt{({\bf r}-{\bf r}')^{2}+\epsilon_{0}^{2}}}d^{3}{\bf r}'.\label{a8}
\end{equation}
The microcanonical ensemble is the correct description of stellar
systems which form an isolated Hamiltonian system in a first
approximation. We can also consider the case of self-gravitating
systems in contact with a thermal bath of non-gravitational origin
which imposes its temperature $T$. For such systems, the correct
description is the canonical ensemble and the statistical equilibrium
state is obtained by minimizing the Boltzmann free energy $F=E-TS$ at
fixed mass. The canonical ensemble is also the correct description of
a gas of self-gravitating Brownian particles (Chavanis, Rosier \& Sire
2002). In this model, the friction and the stochastic fluctuations can
mimick the influence of an external medium (thermostat) to which the
system of origin is coupled.

\subsection{Thermodynamic limit of self-gravitating systems}
\label{sec_tl}

We introduce dimensionless variables such that ${\bf r}=R{\bf r}'$,
${\bf v}=U{\bf v}'$ and $f=(M/R^{3}U^{3})f'$ where $R$ is the box
radius, $M$ is the mass of the system and $U\equiv (GM/R)^{1/2}$ is a
typical velocity obtained by a Virial type argument (or dimensional
analysis). For self-gravitating fermions, the entropy (\ref{fd10}) can
be expressed as
\begin{eqnarray}
S=-Nk_B\mu\int  d^{3}{\bf r}'d^{3}{\bf v}'\qquad\qquad\qquad\qquad\nonumber\\
{\times} \biggl\lbrace {f'\over \mu}\ln\biggl ({f'\over
\mu}\biggr )+\biggl (1-{f'\over \mu}\biggr )\ln\biggl (1-{f'\over
\mu}\biggr )\biggr\rbrace ,
\label{tl1}
\end{eqnarray}
where $\mu=\eta_0\sqrt{G^3 M R^3}$ is the degeneracy parameter (Chavanis
\& Sommeria 1998). Writing
$\mu=(R/R_{*})^{3/2}$ with $R_{*}=h^{2}/GM^{1/3}m^{8/3}$, we note that
the degeneracy parameter is the ratio, to the power $3/2$, of the
system's radius divided by the radius $R_{*}$ of a ``white dwarf
star'' (i.e. a completely degenerate ball of fermions)
with mass $M$. The conservation of mass is equivalent to
\begin{equation}
\int {f'}  d^{3}{\bf r}'d^{3}{\bf v}'=1, \label{tl2}
\end{equation}
and the conservation of energy is equivalent to
\begin{eqnarray}
{ER\over GM^{2}}=\int f' {v^{'2}\over 2}d^{3}{\bf
r}'d^{3}{\bf v}'-{1\over 2}\int {\rho'({\bf r'}_{1})\rho'({\bf
r'}_{2})\over |{\bf r'}_{1}-{\bf r'}_{2}|}d^{3}{\bf r}'_{1}d^{3}{\bf
r}'_{2}.\nonumber\\
\label{tl3}
\end{eqnarray}
Finally, the Massieu function can be written
\begin{equation}
J=N(s[f']+\eta \Lambda[f']),\label{tl4}
\end{equation}
where $s=S/N$, $\eta=\beta GMm/R$ and $\Lambda=-ER/GM^2$. We define
the thermodynamic limit as $N\rightarrow +\infty$ such that
$\mu=\eta_0\sqrt{G^3 M R^3}$, $\Lambda=-ER/GM^2$ and $\eta=\beta
GMm/R$ are fixed. Coming back to physical quantities, it makes sense
to fix $h$, $m$ and $G$. Then, we have the scalings $R\sim N^{-1/3}$,
$T\sim N^{4/3}$, $E\sim N^{7/3}$, $S\sim N$ and $J\sim N$ as
$N\rightarrow +\infty$ (the free energy $F$ scales as $N^{7/3}$). This
is the quantum thermodynamic limit (QTL) for the self-gravitating gas
(Chavanis 2002b, Chavanis \& Rieutord 2003). This thermodynamic limit
is relevant for compact objects with small radii $R\sim N^{-1/3}\ll 1$
such as white dwarfs, neutron stars, fermion balls etc. The usual
thermodynamic limit $N,R\rightarrow +\infty$ with $N/R^{3}$ constant
is clearly not relevant for inhomogeneous systems whose energy is
non-additive (Padmanabhan 1990).

For classical particles with soften potential, the entropy (\ref{a4})
can be expressed as
\begin{equation}
S=-N\int {f'}\biggl\lbrack \ln\biggl ({f'\over N\nu}\biggr
)-1\biggr\rbrack  d^{3}{\bf r}'d^{3}{\bf v}', \label{tl5}
\end{equation}
where $\nu\equiv {m^{4}\over h^{3}}\sqrt{G^{3}MR^{3}}$ is the
counterpart of the degeneracy parameter. For classical
particles, we see that it does not play any fundamental role in
determining the structure of the system since it just appears as
an additional constant term (independent on $f$) in the entropy.
If we only consider the part of entropy that depends on the distribution function, we get
\begin{equation}
S_{R}=-N\int {f'}\ln f'  d^{3}{\bf r}'d^{3}{\bf v}'. \label{tl6}
\end{equation}
This is the relevant part of the entropy functional considered by
Antonov (1962) and Lynden-Bell \& Wood (1968). We can therefore write
$S[f]=S_{R}[f]+S_{I}$ where $S_{I}$ is the constant part (irrelevant).  The
conservation of mass is equivalent to Eq. (\ref{tl2}) and the
conservation of energy is equivalent to
\begin{eqnarray}
{ER\over GM^2}=\int f' {v^{'2}\over 2}d^{3}{\bf
r}'d^{3}{\bf v}'\qquad\qquad\qquad\nonumber\\
-{1\over 2}\int {\rho'({\bf r'}_{1})\rho'({\bf
r'}_{2})\over\sqrt{({\bf r'}_{1}-{\bf
r'}_{2})^{2}+\epsilon^{2}}}d^{3}{\bf r}'_{1}d^{3}{\bf
r}'_{2}, \label{tl7}
\end{eqnarray}
where $\epsilon=r_{0}/R$. As before, the Massieu function is given by
Eq. (\ref{tl4}). We define the thermodynamic limit as $N\rightarrow
+\infty$ such that $\Lambda=-ER/GM^2$, $\eta=\beta GMm/R$ and
$\epsilon=r_{0}/R$ are fixed. Coming back to physical quantities, it
makes sense to fix $r_0$, $m$ and $G$.  Then, we have the scalings
(Chavanis \& Rieutord 2003) $R\sim 1$, $E\sim N^{2}$, $T\sim N$,
$S_{R}\sim N$ and $J_{R}\sim N$ as $N\rightarrow +\infty$ (the free
energy $F_{R}$ scales as $N^{2}$).  These scalings imply that $\nu\sim
N^{1/2}$ (if we fix $h$). Therefore, the (irrelevant) constant part of
the entropy per particle diverges logarithmically as $S_{I}/N\sim
\ln\nu\sim {1\over 2}\ln N\rightarrow +\infty$. This does not seem to be a
crucial problem since this diverging term does not depend on $f$ and
therefore does not affects the structure of the equilibrium state.
However, in a strict sense, there is no thermodynamic limit for
classical self-gravitating particles with soften potential.  This
contrasts with the case of self-gravitating fermions that possess a
rigorous thermodynamic limit (QTL).

Let us finally consider the case of classical self-gravitating
particles without small-scale cut-off. The entropy is given by the
Boltzmann formula (\ref{a4}). When $r_{0}=0$, we know that the
Boltzmann entropy has no global maximum at fixed mass and energy
(Antonov 1962). However, for sufficiently high energies, it has local
entropy maxima that describe metastable gaseous states. The
thermodynamic limit in that context corresponds to $N\rightarrow
+\infty$ such that $\Lambda=-ER/GM^2$ and $\eta=\beta GMm/R$ are of
order unity.  If we fix $m$, $G$ and $T$, we have the scalings $R\sim
N$, $E\sim N$, $S\sim N$, $J\sim N$ and $F\sim N$ as $N\rightarrow
+\infty$. This is the classical thermodynamic limit (CTL), or dilute
limit, for the self-gravitating gas (de Vega \& Sanchez
2002). Physically, it describes {\it metastable} gaseous states that
are not affected by the small-scale cut-off (Chavanis \& Rieutord
2003). As we shall see, these metastable states have considerably long
lifetimes so that this thermodynamic limit is relevant for classical
objects with large radii $R\sim N\gg 1$ such as globular clusters.

\section{Connexion with statistical mechanics}
\label{sec_connexion}

\subsection{Series of equilibria and metastable states}
\label{sec_seq}

\begin{figure}
\centering
\includegraphics[width=8.5cm]{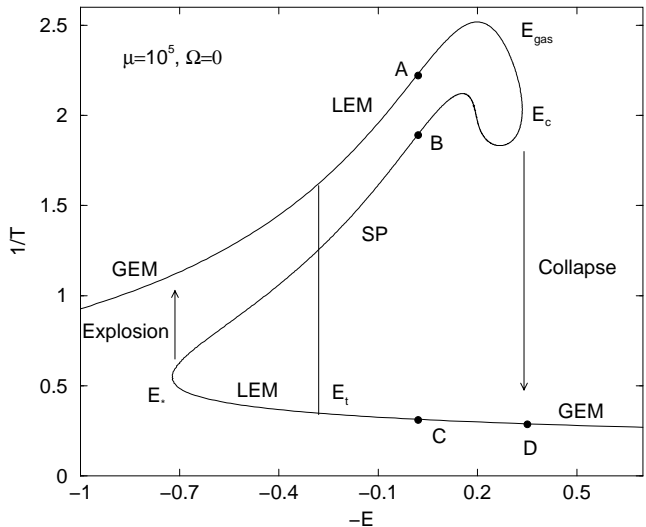}
\caption{Series of equilibria for self-gravitating fermions with small cut-off/large $\mu$/large system size $R$. It has a $Z$-shape structure (dinosaur's neck). There can be several values of inverse temperature  $\beta$ for a given energy $E$. They correspond to local maxima (LEM), global maxima (GEM) or saddle points (SP) of entropy$S[f]$. The same remark applies in the canonical ensemble where the role of $E$ and $\beta$ is reversed.}
\label{le5}
\end{figure}

\begin{figure}
\centering
\includegraphics[width=8.5cm]{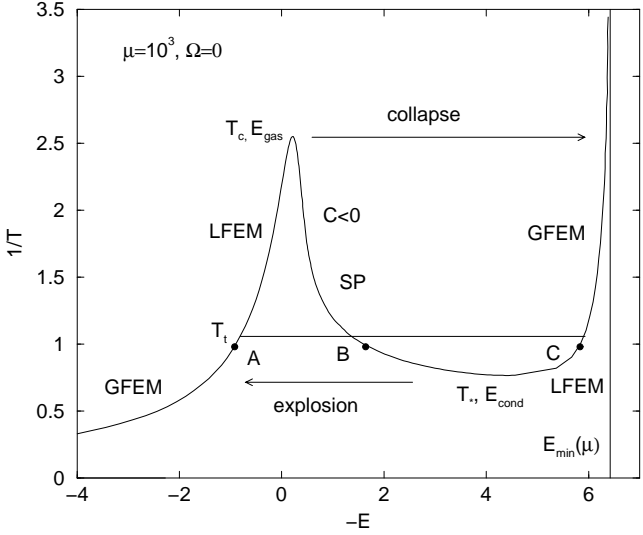}
\caption{Series of equilibria for self-gravitating fermions with large cut-off/small $\mu$/small system size $R$. It has a $N$-shape structure. There is only one  value of inverse temperature
$\beta$ for a given energy $E$. It corresponds to a global maximum of
entropy (GEM). By contrast, there are several values of energy $E$ for
a given $\beta$ in the canonical ensemble. They correspond to local
maxima (LFEM), global maxima (GFEM) or saddle points (SP) of free
energy $J[f]$.}
\label{fel}
\end{figure}

\begin{figure}
\centering
\includegraphics[width=8.5cm]{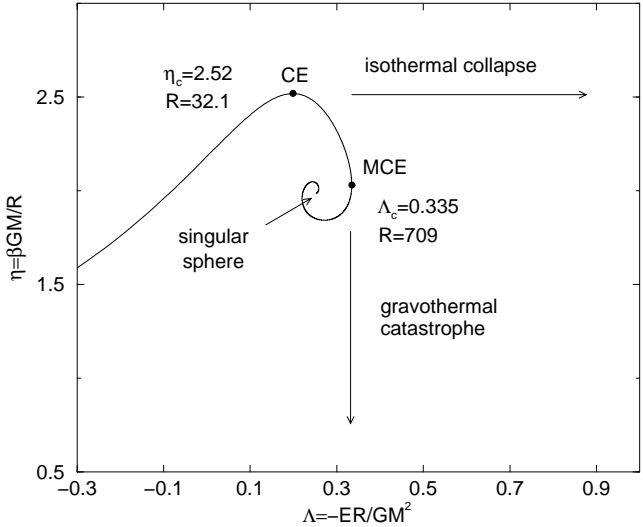}
\caption{Series of equilibria for self-gravitating classical particles without small-scale cut-off. There exists only local entropy maxima (metastable state) or unstable saddle points of entropy. The same remarks hold in the canonical ensemble.  }
\label{etalambda}
\end{figure}

The critical points of entropy $S[f]$ at fixed $E$ and $M$ (i.e., the
distribution functions $f({\bf r},{\bf v})$ which cancel the first
order variations of $S$ at fixed $E$, $M$) form a series of equilibria
parameterized, for example, by the density contrast ${\cal
R}=\rho(0)/\rho(R)$ between the center and the edge of the system (see
Chavanis 2002b). At each point in the series of equilibria corresponds
a temperature $\beta$ and an energy $E$.  In this approach, $\beta$ is
the Lagrange multiplier associated with the conservation of energy in
the variational problem (\ref{fd13}). It has also the interpretation
of a kinetic temperature in the Fermi-Dirac distribution
(\ref{fd14}). We can thus plot $\beta(E)$ along the series of
equilibria. The form of this ``caloric curve'' depends on the value of
the degeneracy parameter $\mu$ in the case of fermions (Chavanis
2002b) and on the soften radius $\epsilon$ for regularized classical
systems (Chavanis \& Ispolatov 2002). It also depends on the dimension
of space $D$ (Sire \& Chavanis 2002, Chavanis 2004a). In $D=3$, the
caloric curve has a $Z$-shape (see Fig. \ref{le5}) for small cut-off
and a $N$-shape (see Fig. \ref{fel}) for large cut-off (for no
cut-off, we recover the well-known spiral of Fig. \ref{etalambda}).
There can be several values of temperature $\beta$ for the same energy
$E$ because the variational problem (\ref{fd12}) can have several
solutions: a local entropy maximum (metastable state), a global
entropy maximum, and one or several saddle points. We must represent
all these solutions on the caloric curve because local entropy maxima
(metastable states) are in general more physical than global entropy
maxima for the timescales achieved in astrophysics. Indeed, the system
can remain frozen in a metastable gaseous phase for a very long
time. This is the case, in particular, for globular clusters and for
the gaseous phase of fermionic matter (at high energy and high
temperature). The time required for a system placed in a metastable
gaseous state to collapse is in general tremendously long and
increases exponentially with the number $N$ of particles. Thus,
$t_{life}\rightarrow +\infty$ in the thermodynamic limit $N\rightarrow
+\infty$. The robustness of metastable states is due to the long-range
nature of the gravitational potential. Therefore, at high temperatures
and high energies, the global entropy maximum is not physically
relevant.  Condensed objects (e.g., planets, stars, white dwarfs,
fermion balls,...) only form below a critical energy $E_{c}$ (Antonov
energy) or below a critical temperature $T_{c}$ (Jeans temperature),
when the gaseous metastable phase ceases to exist. The point where the
metastable phase disappears is called a spinodal point.

\subsection{Microcanonical ensemble} \label{sec_micro}

Let us explain things differently so as to make a closer contact with
statistical mechanics. In statistical mechanics, one usually starts
with the density of states
\begin{eqnarray}
g(E)=\int \delta \lbrack E-H({\bf r}_1,...,{\bf r}_N,{\bf
v}_1,...,{\bf v}_N)\rbrack\prod_{i=1}^N {d^{3}{\bf r}_i d^3 {\bf
v}_i\over ({h\over m})^{3N}},\nonumber\\
 \label{m1}
\end{eqnarray}
where $H$ is the Hamiltonian. For our system
\begin{equation}
H={1\over 2}\sum_{i=1}^{N}mv_i^2-\sum_{i<j}{Gm^2\over |{\bf
r}_i-{\bf r}_j|}.\label{m2}
\end{equation}
The density of states is the normalization factor of the $N$-body microcanonical distribution
\begin{equation}
P_{N}({\bf r}_{1},{\bf v}_{1},...,{\bf r}_{N},{\bf v}_{N})={1\over g(E)}\delta  (E-H),
\label{m2a}
\end{equation}
stating that all accessible microstates are equiprobable.

\begin{figure}
\centering
\includegraphics[width=8.5cm]{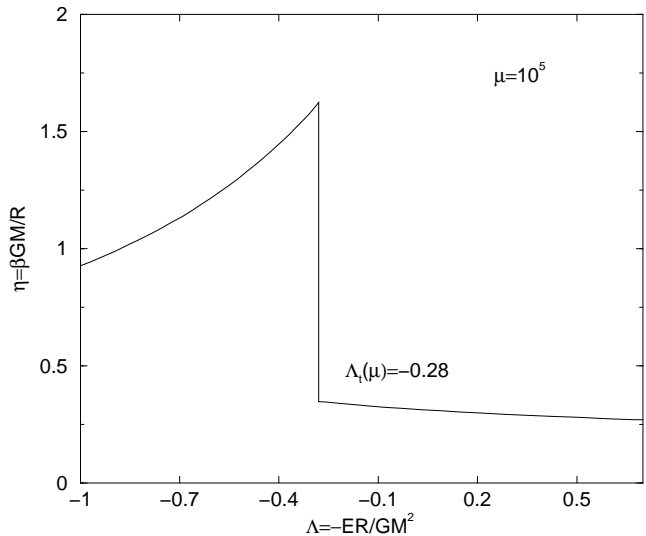}
\caption{Strict microcanonical caloric curve for $\mu=10^{5}$ (the control parameter is the energy $E$). This figure is obtained from Fig. \ref{le5} by keeping only global entropy maxima. It corresponds therefore to the true caloric curve $\beta_{micro}(E)$ which  is univalued (Gross 2003). For $N\rightarrow +\infty$, there is a discontinuity of temperature at the transition energy $E_{t}(\mu)$. For finite $N$ systems, this discontinuity is smoothed out. Although this caloric curve is correct in a strict sense, it is {\it not} physical because it ignores metastable states that have an infinite lifetime in the thermodynamic limit. The physical caloric curve is obtained from Fig. \ref{le5} by discarding the unstable saddle points of entropy that form the intermediate branch (see Fig. \ref{el5phys}).}
\label{caloric}
\end{figure}

Introducing the probability $W(\lbrace n_i\rbrace)$ of the state
$\lbrace n_i\rbrace$, we can rewrite the density of states in the
form
\begin{equation}
g(E)=\sum_{E(\lbrace n_{i}\rbrace )=E}W(\lbrace n_i\rbrace), \label{m4}
\end{equation}
where the sum runs over all macrostates with energy $E$.
Introducing the entropy $S=\ln W$ of the state $\lbrace
n_i\rbrace$ and taking the continuum limit, the density of states
can be expressed formally as
\begin{equation}
g(E)=\int {\cal D}f e^{S[f]}\delta (E-E[f])\delta (M-M[f]),
\label{m5}
\end{equation}
where the sum runs over all distribution functions and $S[f]$ is the
Fermi-Dirac entropy (\ref{fd10}) if the particles are fermions and the
Boltzmann entropy (\ref{a4}) for classical particles (in that case,
the gravitational potential must be regularized otherwise the density
of states (\ref{m1}) diverges). We now define the microcanonical
entropy by $S_{micro}(E)=\ln g(E)$ and the microcanonical temperature
by $\beta_{micro}=dS_{micro}(E)/dE$. By definition, the caloric curve
$\beta_{micro}(E)$ is uni-valued (Gross 2003). In the thermodynamic
limit defined in Sec. \ref{sec_tl}, the entropy $S[f]$ scales as $\sim
N$, that is $S[f]=Ns[f]$ where $s\sim 1$ is the entropy per
particle. Therefore, for $N\rightarrow +\infty$, the integral in
Eq. (\ref{m5}) is dominated by the state $f_{global}({\bf r},{\bf v})$
which is the global maximum of $S[f]$ at fixed $M$ and $E$ (rigorously
speaking, we should work with the dimensionless quantities defined in
Sec. \ref{sec_tl} to get rid of the $N\rightarrow +\infty$
dependence). Then, $g(E)\simeq e^{S[f_{global}]}$, $S_{micro}\simeq
S[f_{global}]$ and $\beta_{micro}=\delta S/\delta E=\beta$. However,
this approach fails to take into account metastable states (local
maxima of $S[f]$ at fixed $M$ and $E$), which are of considerable
interest in astrophysics. Indeed, equilibrium statistical mechanics
tells nothing about timescales; a kinetic theory is required in that
case.  As explained above, these metastable states can persist for
very long times. They correspond to the observed ``diluted''
structures in the universe (e.g., globular clusters). Therefore, the
caloric curve $\beta_{micro}(E)$ does not describe the system
adequately. The series of equilibria $\beta(E)$ contain more
information as they show local and global entropy maxima (as well as
unstable saddle points).  The curve $\beta_{micro}(E)$ can be deduced
from $\beta(E)$ by keeping only global entropy maxima (see
Fig. \ref{caloric}). If we adopt this procedure, we find that the
system exhibits a first order {\it microcanonical} phase transition
(provided that the system size $\mu$ is sufficiently large) at a
transition energy $E_t(\mu)$ where the gaseous phase and the condensed
phase have the same entropy (Chavanis 2002b). In the strict
thermodynamic limit $N\rightarrow +\infty$, this phase transition is
marked by a discontinuity of temperature. In fact, for finite $N$
systems, the mean-field approximation $g(E)\simeq e^{S[f_{global}]}$
breaks down near the transition energy (Chavanis \& Ispolatov
2002). This is because the contribution of the local entropy maximum
in the functional integral (\ref{m5}) becomes more and more important
as we approach $E_{t}$. For the saddle point approximation to be
valid, the number of particles must scale as $N\sim
|\Lambda-\Lambda_{t}|^{-1}$ for $\Lambda\rightarrow
\Lambda_{t}$ (see Sec. \ref{sec_tt}).  Thus, for large but {finite} $N$, the
temperature variation is sharp but remains {\it continuous} at the
transition. We again emphasize that, due to the existence of
metastable states, this phase transition may not be physically
relevant. The true phase transition (gravothermal catastrophe) will
rather take place at, or near, the spinodal point $E_c$ (Antonov
energy) where the metastable branch disappears. Estimating the
lifetime of a metastable state by the Kramers formula $t_{life}\sim
e^{\Delta S}$, where $\Delta S$ is the height of the entropic barrier
(difference of entropy between the local maximum and the saddle
point), we find that $t_{life} \sim {\rm
exp}\lbrack{2\lambda'N(\Lambda_{c}-\Lambda)^{3/2}}\rbrack$ with
$\lambda'\simeq 0.863159...$ (see Sec. \ref{sec_persistence}). Except
in the vicinity of the critical point $\Lambda_{c}$, the lifetime of a
gaseous metastable state is enormous as it increases exponentially
with the number of particles.  Thus, metastable states have a
considerable interest in astrophysics.

\begin{figure}
\centering
\includegraphics[width=8.5cm]{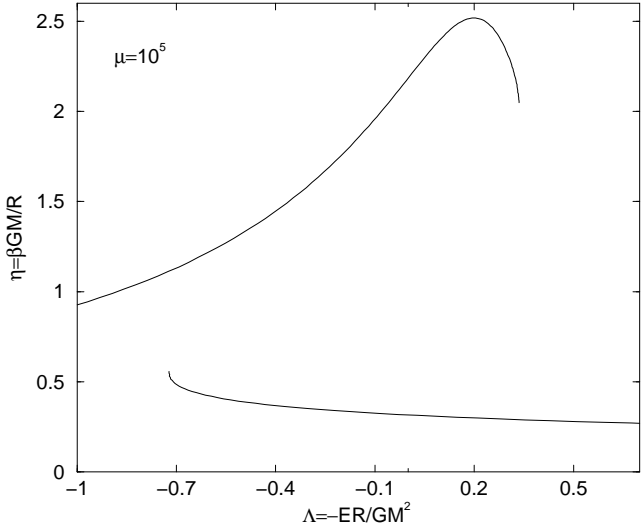}
\caption{Physical microcanonical caloric curve for $\mu=10^{5}$  (the control parameter is the energy $E$). This figure displays stable and metastable equilibrium states. The metastable states have infinite lifetime in the thermodynamic limit so they are physically relevant. The first order microcanonical phase transition of Fig. \ref{caloric} does not take place in practice. Due to the existence of metastable states, the system displays a microcanonical hysteretic cycle marked by a ``collapse'' and an ``explosion'' at the spinodal points where the branch of metastable states disapears (Chavanis \& Rieutord 2003).}
\label{el5phys}
\end{figure}

If we now consider the case of classical particles ($\hbar\rightarrow
0$ or $\mu\rightarrow +\infty$), the transition energy $E_{t}(\mu)$ is
rejected to $+\infty$ so that the whole branch of gaseous states is
metastable. This ``no cut-off'' limit is relevant to classical objects
such as globular clusters or to the interstellar medium because, for
these systems, the size of the particles (stars and atoms) clearly
does not matter.  In that case, the series of equilibria $\beta(E)$
forms a spiral (see Fig. \ref{etalambda}) indicating the existence of
one local entropy maximum and one (or several) saddle points of
entropy for a given energy (Lynden-Bell \& Wood 1968). This spiral is
the limiting form, for $\mu\rightarrow +\infty$, of the fermionic
caloric curve (see Fig. 11 in Chavanis 2002b). In this limit, the
branch of ``collapsed'' states (condensed phase) coincides with the
$x$-axis where $\beta=0$. It corresponds to configurations made of two
particles in contact ($\sim$ binary star) surrounded by a hot halo with
$T\rightarrow +\infty$. This ``binary+halo'' configuration has an
infinite entropy so, in a sense, it is the most probable configuration
in the microcanonical ensemble (see Appendix A of Sire \& Chavanis
2002). However, for sufficiently large energies (above the Antonov
point), these configurations must be discarded ``by hands'' because
they are reached for inaccessibly large times. Therefore, for
classical particles, the physical caloric curve $\beta_{physical}(E)$
is obtained by taking $g(E)\simeq e^{S[f_{local}]}$ and
$S_{physical}\simeq S[f_{local}]$ where $f_{local}$ is the local
entropy maximum at fixed mass and energy (Chavanis 2003).

\subsection{Canonical ensemble} \label{sec_cano}

\begin{figure}
\centering
\includegraphics[width=8.5cm]{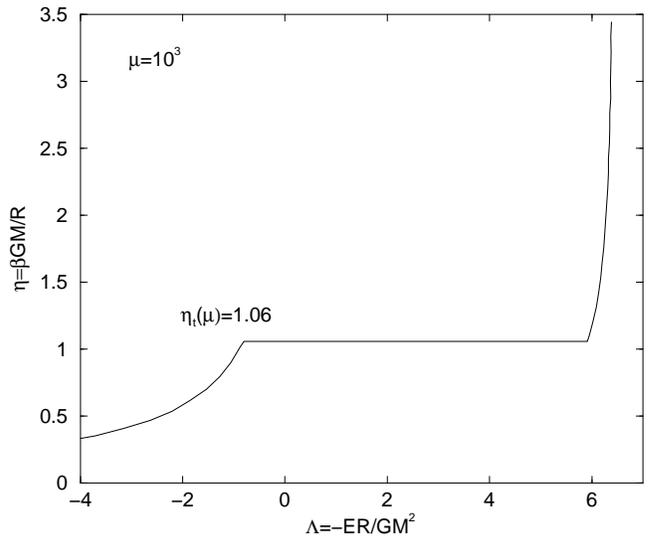}
\caption{Strict canonical caloric curve for $\mu=10^{3}$ (the control parameter is the inverse temperature $\beta$). This figure is obtained from Fig. \ref{fel} by keeping only global maxima of free energy. It corresponds therefore to the true canonical caloric curve which  is univalued and does not display negative specific heats contrary to the corresponding microcanonical caloric curve (see Fig. \ref{el3micro}). For $N\rightarrow +\infty$, there is a discontinuity of energy (latent heat) at the transition temperature $T_{t}(\mu)$. For finite $N$ systems, this discontinuity is smoothed-out.  Although this canonical caloric curve is correct in a strict sense, it is {\it not} physical because it ignores metastable states that have an infinite lifetime in the thermodynamic limit $N\rightarrow +\infty$. The physical canonical caloric curve is obtained from Fig. \ref{fel} by discarding the unstable saddle points of free energy that form the intermediate branch with negative specific heats (see Fig. \ref{el3phys}).}
\label{caloricCan}
\end{figure}

\begin{figure}
\centering
\includegraphics[width=8.5cm]{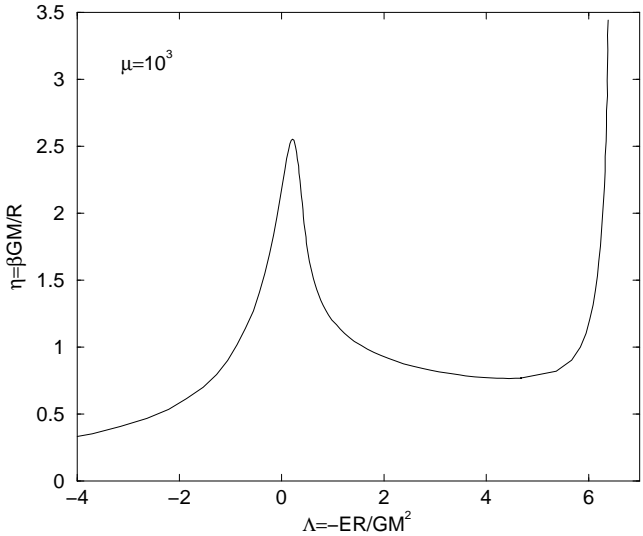}
\caption{Microcanonical caloric curve for the same cut-off/degeneracy parameter $\mu$/system's radius $R$ as Fig. \ref{caloricCan} (the control parameter is the energy $E$). It has a $N$-shape structure and displays a stable region with negative specific heats. All the equilibria are global maxima of entropy. In the canonical ensemble the region of negative specific heats is replaced by a phase transition (see Fig. \ref{caloricCan}). The curves of Figs. \ref{caloricCan} and \ref{el3micro} are similar to those obtained by Padmanabhan (1990) with his toy model consisting in $N=2$ stars in gravitational interaction.}
\label{el3micro}
\end{figure}

\begin{figure}
\centering
\includegraphics[width=8.5cm]{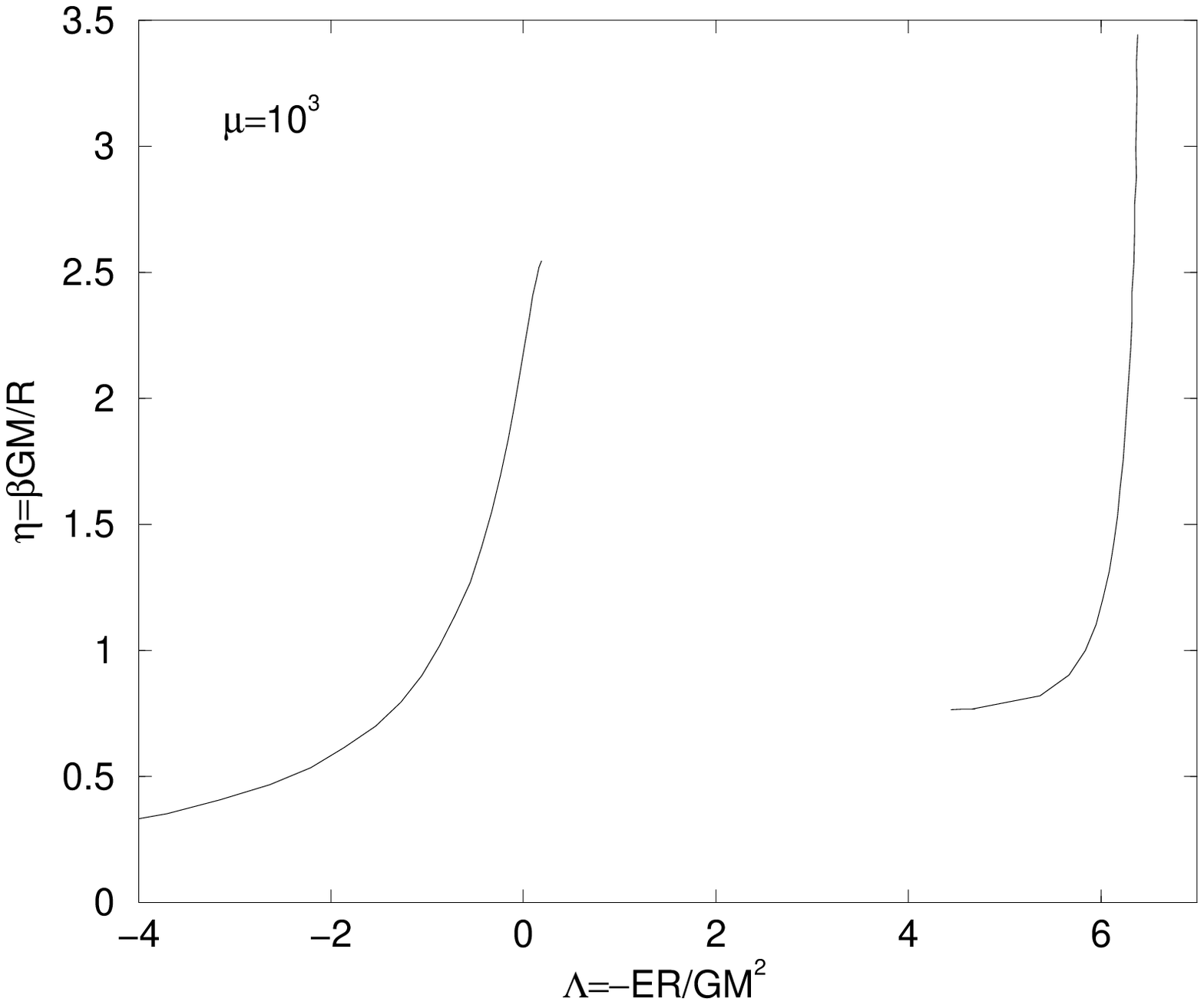}
\caption{Physical canonical caloric curve for $\mu=10^{3}$ (the control parameter is the inverse temperature $\beta$). This figure displays stable and metastable equilibrium states. The metastable states have infinite lifetime in the thermodynamic limit so they are physically relevant. The first order canonical phase transition of Fig. \ref{caloricCan} does not take place in practice. Due to the existence of metastable states the system displays a canonical hysteretic cycle marked by a ``collapse'' and an ``explosion'' at the spinodal points where the branch of metastable states disapears (Chavanis \& Rieutord 2003).}
\label{el3phys}
\end{figure}

In the canonical ensemble, the object of fundamental importance is the
partition function
\begin{equation}
Z(\beta)=\int e^{-\beta H({\bf r}_1,...,{\bf r}_N,{\bf
v}_1,...,{\bf v}_N)}\prod_{i=1}^N {d^{D}{\bf r}_i d^D {\bf v}_i\over ({h\over m})^{DN}}.
\label{can1}
\end{equation}
which is the normalization of the $N$-body canonical distribution
\begin{equation}
P_{N}({\bf r}_1,{\bf
v}_1,...,{\bf r}_N,{\bf
v}_N)={1\over Z(\beta)} e^{-\beta H}.
\label{can1a}
\end{equation}
The partition function  can be rewritten
\begin{eqnarray}
Z(\beta)=\sum_{\lbrace n_{i}\rbrace } e^{-\beta E(\lbrace
n_{i}\rbrace )} W(\lbrace n_{i}\rbrace )\qquad\qquad\nonumber\\
=\sum_{\lbrace
n_{i}\rbrace } e^{S(\lbrace n_{i}\rbrace )-\beta E(\lbrace
n_{i}\rbrace )} =\sum_{\lbrace n_{i}\rbrace } e^{J(\lbrace
n_{i}\rbrace )}, \label{can2}
\end{eqnarray}
where $J(\lbrace n_{i}\rbrace )=S(\lbrace n_{i}\rbrace )-\beta
E(\lbrace n_{i}\rbrace )$ is the ``free energy'' of the macrostate
$\lbrace n_{i}\rbrace$ and the sum runs over all macrostates (in the
present context, $J[f]=S[f]-\beta E[f]$ is a more natural functional
than the usual free energy $F[f]=E[f]-TS[f]$).  Taking the continuum
limit, the partition function can be expressed formally as
\begin{equation}
Z(\beta)=\int {\cal D}f e^{J[f]}\delta (M-M[f]), \label{cano1}
\end{equation}
where the sum runs over all distribution functions and
$J[f]$ is the Fermi-Dirac free energy if the
particles are fermions and the Boltzmann free energy for classical
particles (in that case, the gravitational potential must be
regularized otherwise the partition function (\ref{can1})
diverges). Note that Eq. (\ref{cano1}) can also be obtained from the exact
formula
\begin{equation}
Z(\beta)=\int e^{-\beta E}g(E)dE, \label{cano1bis}
\end{equation}
by substituting Eq. (\ref{m5}) for $g(E)$ and carrying out the
integration over $E$. We now define the canonical free energy by
$F_{cano}=-(1/\beta)\ln Z$.  The average energy of the system at
temperature $T$ can be written $\langle E\rangle_{cano}=-\partial\ln
Z/\partial\beta$.  By definition, the caloric curve $\langle
E\rangle_{cano}(\beta)$ is uni-valued.  In the thermodynamic limit
defined in Sec. \ref{sec_tl}, the free energy $J[f]$ scales as
$\sim N$. Therefore, for $N\rightarrow +\infty$, the integral in
Eq. (\ref{cano1}) is dominated by the state $f_{global}({\bf r},{\bf
v})$ which is the global maximum of $J[f]$ at fixed $M$. Then,
$Z(\beta)\simeq e^{J[f_{global}]}$, $F_{cano}\simeq
-(1/\beta)J[f_{global}]=E[f_{global}]-TS[f_{global}]$ and $\langle
E\rangle_{cano}=-\delta J/\delta\beta=E$. Metastable states (local
maxima of $J[f]$ at fixed $M$) can be taken into account by plotting
the full curve $E(\beta)$. It is obtained from $\beta(E)$ defined in
Sec. \ref{sec_seq} by simply reversing the graph since the critical
points of the variational problems (\ref{fd12}) and (\ref{fd15}) are
the same (see Sec. \ref{sec_fd}). The curve $\langle
E\rangle_{cano}(\beta)$ can be deduced from $E(\beta)$ by keeping only
global maxima of free energy (see Fig. \ref{caloricCan}). If we adopt
this procedure, we find that the system exhibits a first order {\it
canonical} phase transition at a transition temperature $T_t(\mu)$
where the gaseous phase and the condensed phase have the same free
energy (Chavanis 2002b). This phase transition is marked by a
discontinuity of energy (latent heat). In fact, the mean-field
approximation $Z(\beta)\simeq e^{J[f_{global}]}$ breaks down near the
transition temperature. For large but finite $N$, the energy variation
is sharp but remains {\it continuous} at the transition. For the
saddle point approximation to be valid, the number of particles must
scale as $N\sim |\eta-\eta_{t}|^{-1}$ for $\eta\rightarrow \eta_{t}$
(see Sec. \ref{sec_tt}). We again emphasize that, due to the existence
of metastable states, this phase transition may not be relevant and
that the physical phase transition (isothermal collapse) takes place
at, or near, the spinodal point $T_c$ (Jeans temperature) where the
metastable branch disappears (see Fig. \ref{el3phys}). Estimating the
lifetime of a metastable state by the Kramers formula $t_{life}\sim
e^{\Delta F/k_{B}T}$, where $\Delta F$ is the height of the potential
barrier (difference of free energy between the local maximum and the
saddle point), we find that $t_{life} \sim {\rm exp}\lbrace 2\lambda N
(\eta_{c}-\eta)^{3/2}\rbrace$ with $\lambda\simeq 0.16979815...$ (see
Sec.
\ref{sec_persistence}). Except in the vicinity of the critical point
$\eta_{c}$, the lifetime of a gaseous metastable state is enormous as
it increases exponentially with the number of particles. Metastable
states are therefore highly robust. For classical objects
($\hbar\rightarrow 0$), the transition temperature $T_{t}(\mu)$ is
rejected to $+\infty$ so that the whole branch of gaseous states is
metastable.  In that case, the series of equilibria $E(\beta)$ forms a
spiral (see Fig. \ref{etalambda}) indicating the existence of one
local minimum of free energy $F$ and one (or several) saddle points of
free energy for a given temperature. In the classical limit, the
branch of ``collapsed'' states (condensed phase) is rejected to
$E\rightarrow
-\infty$.  It corresponds to configurations where all the particles
have collapsed at $r=0$. This ``Dirac peak'' configuration has an
infinite free energy $F=-\infty$ (due to the infinite binding energy)
so, in a sense, it is the most probable configuration in the canonical
ensemble (see Appendix B of Sire \& Chavanis 2002). This differs from
the binary star surrounded by a hot halo in the microcanonical
ensemble.  However, for sufficiently large temperatures (above the
Emden-Jeans point), these configurations must be discarded ``by
hands'' because they are reached for inaccessibly large
times. Therefore, for classical particles, the physical caloric curve
$E_{physical}(\beta)$ is obtained by taking $Z(\beta)\simeq
e^{J[f_{local}]}$ and $J_{physical}\simeq J[f_{local}]$ where
$f_{local}$ is the local maximum of free energy $J$ at fixed mass and
temperature (Chavanis 2003).

\subsection{Grand canonical ensemble} \label{sec_gc}

In the grand canonical ensemble, the partition function is
\begin{eqnarray}
Z_{GC}(\beta,\mu)=\sum_{N=0}^{+\infty} e^{N\mu\over k_{B}T} Z_N(\beta).
\label{gdZ}
\end{eqnarray}
Using Eq. (\ref{cano1}), we get
\begin{eqnarray}
Z_{GC}=\sum_{N=0}^{+\infty}\int {\cal D}f e^{J[f]}e^{N\mu\over k_{B}T}\delta (N-N[f])\nonumber\\
= \int {\cal D}f e^{J[f]+\beta \mu N[f]}=\int {\cal D}f e^{G[f]},
\label{gdZ2}
\end{eqnarray}
where $G[f]=J[f]+\beta \mu N[f]$ is the grand potential. Of course,
the expression (\ref{cano1}) for $Z_N$ is correct only for $N\gg
1$. However, the contribution of small $N$ terms in the grand
partition function (\ref{gdZ}) is expected to be weak so that
Eq. (\ref{gdZ2}) provides a good approximation of the series. Now, the
grand potential $G[f]$ scales as $\sim N_{0}\equiv {R\over \beta
Gm^{2}}$. Therefore, in the thermodynamic limit $R\rightarrow +\infty$
with fixed $\beta$, $G$ (gravitational constant) and $m$, the
partition function $Z_{GC}$ is dominated by the distribution $f$ which
maximizes the grand potential $G[f]$ at fixed $\beta$ and $\mu$.  This
problem has been considered  for classical particles in
$D=3$ (Chavanis 2003). We shall reserve for a future work the study of
self-gravitating fermions in the grand canonical ensemble.

\section{First order microcanonical and canonical phase transitions}
\label{sec_corr}

\subsection{Maxwell constructions and critical points}
\label{sec_maxwell}

\begin{figure}
\centering
\includegraphics[width=8.5cm]{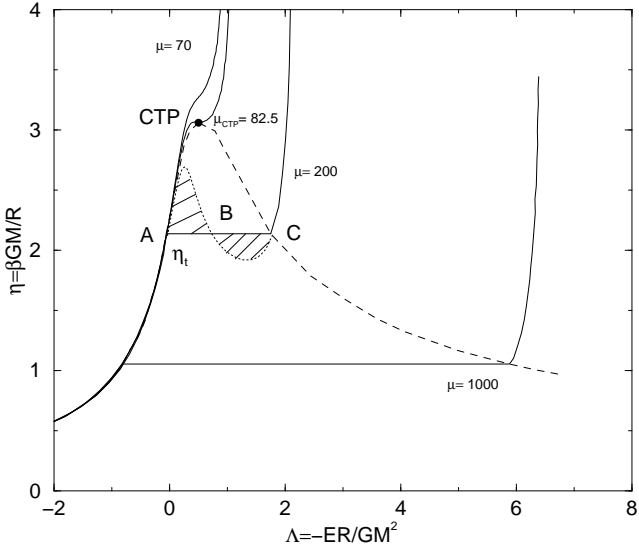}
\caption{Horizontal Maxwell plateau associated with a canonical first order phase transition.}
\label{tricritique}
\end{figure}

\begin{figure}
\centering
\includegraphics[width=8.5cm]{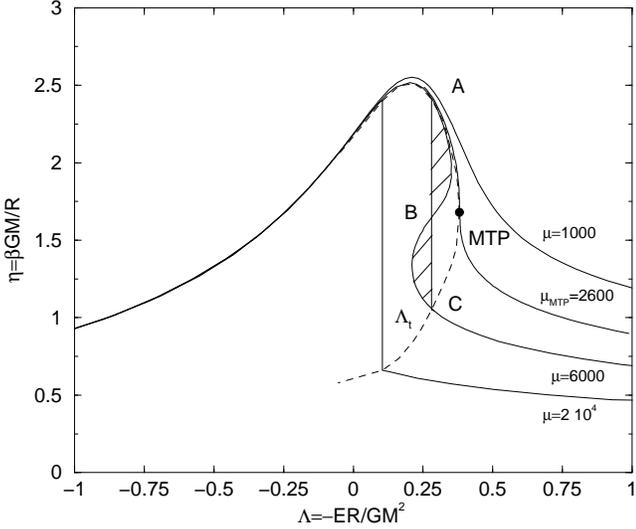}
\caption{Vertical Maxwell plateau associated with a microcanonical first order phase transition. }
\label{tricritiquemicro}
\end{figure}

The deformation of the caloric curve when we vary the degeneracy
parameter $\mu$/ system size $R$ is represented in
Figs. \ref{tricritique} and
\ref{tricritiquemicro}. Similar curves are obtained for a hard
sphere gas or a soften potential (Chavanis \& Ispolatov 2002) instead
of fermions. In that case, $1/\mu$ plays the
role of the cut-off radius $a$ or soften radius $r_{0}$.

For $\mu<\mu_{CTP}\simeq 82.5$, the curve $\beta(E)$ defining the
series of equilibria is monotonic, so there is no phase
transition. For $\mu>\mu_{CTP}$, the curve $E(\beta)$ is multivalued
so that a canonical first order phase transition is expected. The
temperature of transition in the canonical ensemble can be obtained by
a Maxwell construction as for the familiar Van der Waals gas. The
equal area Maxwell condition $A_{1}=A_{2}$ (see
Fig. \ref{tricritique}) can be expressed as
\begin{equation}
\int_{E_{A}}^{E_{C}}(\beta-\beta_{t})dE=0, \label{max1}
\end{equation}
where $E_A$ is the energy of the gaseous phase and $E_B$ the
energy of the condensed phase at the transition temperature
$T_{t}$ . Since $dS=\beta dE$, one has
\begin{equation}
S_{C}-S_{A}-\beta_{t}(E_{C}-E_{A})=0. \label{max2}
\end{equation}
Introducing the free energy $J=S-\beta E$, we verify that
the Maxwell construction is equivalent to the equality of the free
energy of the two phases at the transition:
\begin{equation}
J_{A}=J_{C}. \label{max3}
\end{equation}
If we keep only global maxima of free energy as in
Fig. \ref{caloricCan}, the winding branch has to be replaced by a
horizontal plateau. We see on Fig. \ref{tricritique} that the extent
of the plateau decreases as $\mu$ decreases.  At the canonical
critical point $\mu_{CTP}$, the plateau disappears and the curve
presents an inflexion point.

For $\mu>\mu_{MTP}\simeq 2600$, the curve $\beta(E)$ is multivalued so
that a microcanonical first order phase transition is expected to
occur (in addition to the canonical first order phase transition
described previously). The energy of transition can be obtained by a
vertical Maxwell construction. The equal area Maxwell condition
$A_{1}=A_{2}$ (see Fig. \ref{tricritiquemicro}) can be expressed as
\begin{equation}
\int_{\beta_{A}}^{\beta_{C}}(E-E_{t})d\beta=0, \label{max4}
\end{equation}
where $T_A$ and $T_C$ are the temperatures of
the two phases at the transition energy  $E_{t}$. Since $dJ=-E
d\beta$, one has
\begin{equation}
J_{C}-J_{A}+E_{t}(\beta_{C}-\beta_{A})=0. \label{max5}
\end{equation}
Thus, the Maxwell construction is equivalent to the equality of the
entropy of the two phases at the transition:
\begin{equation}
S_{A}=S_{C}. \label{max6}
\end{equation}
If we keep only global maxima of entropy as in Fig. \ref{caloric}, the
winding branch has to be replaced by a vertical plateau. We see that
the extent of the plateau decreases as $\mu$ decreases. At the
microcanonical critical point $\mu_{MTP}$, the plateau disappears and
the curve presents an inflexion point.

Therefore, for $\mu>\mu_{MTP}$, we expect a microcanonical and a
canonical first order phase transition, for $\mu_{CTP}<\mu<\mu_{MTP}$
only a canonical first order phase transition and for $\mu<\mu_{CTP}$
no phase transition at all. We emphasize, however, that due to the
presence of long-lived metastable states, the first order phase
transitions and the plateau are not relevant for the timescales of
interest (see Sec. \ref{sec_connexion}). Only the zeroth order phase
transitions (gravothermal catastrophe and isothermal collapse) marked
by a discontinuity of entropy or free energy are physically relevant.

\subsection{Validity of the saddle point approximation near the transition point}
\label{sec_tt}

\begin{figure}
\centering
\includegraphics[width=8.5cm]{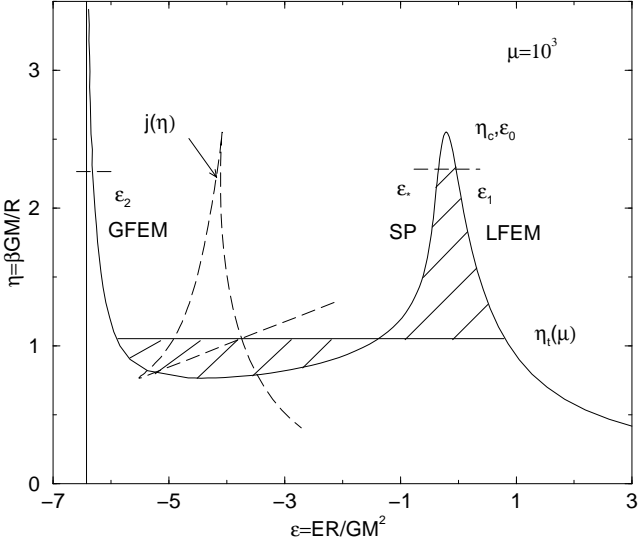}
\caption{Caloric curve of the self-gravitating Fermi gas with
$\mu=1000$. In the canonical ensemble, the temperature of
transition is determined by a Maxwell construction. For
$\eta<\eta_{t}$ ($T>T_{t}$), the gaseous states are global maxima
of free energy $J[f]$ and the condensed states local maxima (see
the dashed curve). The situation is reversed for $\eta>\eta_{t}$
($T<T_{t}$).  } \label{EbetaJMU1e3}
\end{figure}

\begin{figure}
\centering
\includegraphics[width=8.5cm]{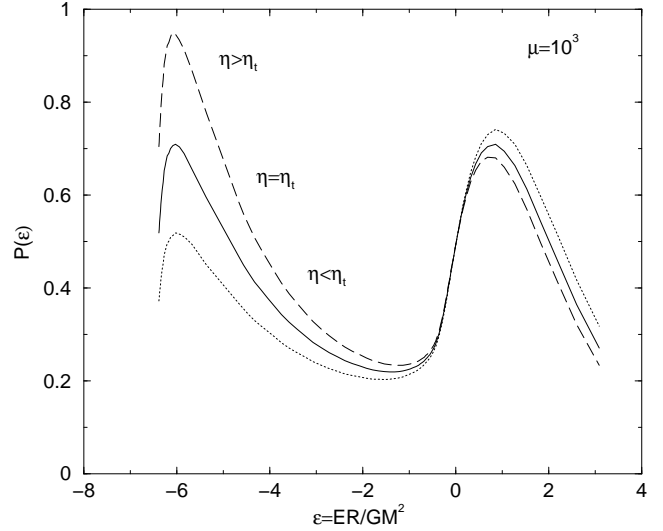}
\caption{Distribution of energies $P(E)\propto {\rm
exp}(S(E)-\beta E)$ in the canonical ensemble for different values
of temperature (here $\mu=1000$). The entropy $S(E)$ has been
calculated in the mean-field approximation, using Eq. (\ref{cc9}).
The distribution of energies has a bimodal structure
characteristic of a first order canonical phase transition. For
$\eta=\eta_{t}=1.052$, the two phases have the same probability.
For $\eta=1<\eta_{t}$, the gaseous phase is the ``most probable'' and
for $\eta=1.1>\eta_{t}$, the condensed phase is the ``most
probable'' (in a strict sense).} \label{ProbE}
\end{figure}

In this section, we discuss the validity of the saddle point
approximation near the transition point. In the canonical
ensemble, the partition function can be written
\begin{equation}
Z(\beta)=\int e^{-\beta E}g(E)dE,
\label{b1}
\end{equation}
where $g(E)$ is the density of states with energy $E$. Introducing the entropy
$S(E)=\ln g(E)$, we can rewrite the partition function as
\begin{eqnarray}
Z(\beta)=\int e^{S(E)-\beta E}dE=\int e^{J(E)}dE=\int
e^{Nj(E)}dE,\nonumber\\
 \label{b2}
\end{eqnarray}
where $J(E)=S(E)-\beta E$ is the ``free energy''. As explained in
Sec. \ref{sec_connexion}, the equilibrium states correspond to maxima
of $J$. The condition $J'(E)=0$, leading to $S'(E)=\beta$, determines
a series of equilibria $E(\beta)$. We shall consider the case where
the series of equilibria $E(\beta)$ has the $N$-shape structure of
Fig. \ref{EbetaJMU1e3}. The transition temperature $T_{t}(\mu)$, is
determined by a Maxwell construction (see Sec
\ref{sec_maxwell}). From Eq. (\ref{b2}), the distribution of
energies at temperature $T$ is given by
\begin{equation}
P(E)={1\over Z(\beta)}e^{J(E)}.
\label{b3}
\end{equation}
It has a bimodal structure as shown in Fig. \ref{ProbE}. The energies
where $P(E)$ is maximum (denoted $E_{1}$ and $E_{2}$) correspond to
the stable (GFEM) and metastable (LFEM) states on Fig. \ref{EbetaJMU1e3}. The energy
where $P(E)$ is minimum (denoted $E_{*}$) corresponds to the unstable
states (SP) on Fig. \ref{EbetaJMU1e3}. For $T>T_{t}$, gaseous
configurations (high energies) are more probable than condensed
configurations (low energies). This is the opposite for
$T<T_{t}$. Note that the notion of ``more probable'' is delicate since
the system can remain blocked in a metastable (``less probable'')
state for very long time, making that state the physically most likely
state.

For $N\rightarrow +\infty$, the partition function can be
approximated by
\begin{equation}
Z(\beta)=e^{Nj(E_{1})}+e^{Nj(E_{2})},
\label{b4}
\end{equation}
where $E_{1}$ and $E_{2}$ are the energies at which $J(E)$ is
maximum. We now wish to obtain the strict caloric curve $\langle
E\rangle_{cano} (\beta)$ defined in Sec. \ref{sec_connexion}.  When
$T$ is not too close from the transition temperature $T_{t}$ and
$N\rightarrow +\infty$, we need just keep the contribution of the
global maximum of free energy as explained previously. To investigate
the situation close to the transition temperature, we rewrite the
partition function (\ref{b4}) as
\begin{equation}
Z(\beta)=e^{Nj(E_{1})}\biggl\lbrack 1+e^{N(j(E_{2})-j(E_{1}))}\biggr\rbrack.
\label{b5}
\end{equation}
Now, close to the transition point, we have
$j(E_{1})=j(E_{2})+\lambda^{2}(\eta-\eta_{t})$, where $\lambda$ is a
constant of order unity depending on $\mu$ (see
Fig. \ref{EbetaJMU1e3}, dashed line). Therefore,
\begin{equation}
Z(\beta)=e^{Nj(E_{1})}\biggl\lbrack 1+e^{-N\lambda^{2}\Delta\eta}\biggr\rbrack,
\label{b6}
\end{equation}
where $\Delta\eta=\eta-\eta_{t}$. Thus, the saddle point
approximation is valid for $N|\eta-\eta_{t}|\gg 1$. This requires
increasing large values of $N$ as we approach the transition
temperature $T_{t}$.

\begin{figure}
\centering
\includegraphics[width=8.5cm]{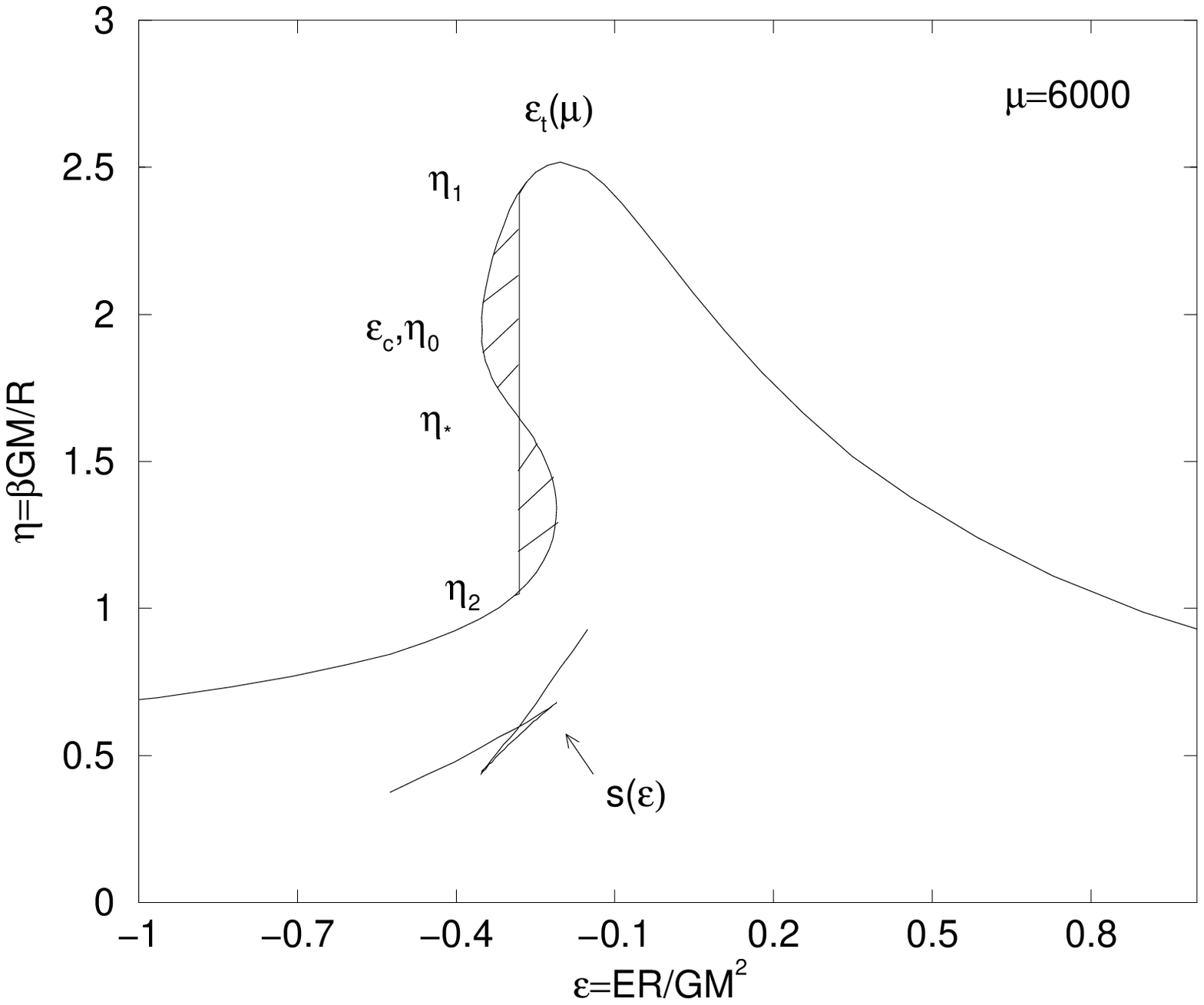}
\caption{Caloric curve of the self-gravitating Fermi gas with
$\mu=10^5$. In the microcanonical ensemble, the energy of
transition is determined by a Maxwell construction. For
$\epsilon>\epsilon_{t}$, the gaseous states are global
maxima of entropy $S[f]$ and the condensed states are local maxima
(see the dashed curve). The situation is reversed for
$\epsilon<\epsilon_{t}$. } \label{entropie}
\end{figure}

In the microcanonical ensemble, the density of states can be
written
\begin{equation}
g(E)=\int {\cal D}f e^{S[f]}\delta (E-E[f])\delta (M-M[f]).
\label{m5bi}
\end{equation}
We shall consider the case where the series of equilibria $\beta(E)$
has the $S$-shape structure of Fig. \ref{entropie}. The transition
energy $E_{t}(\mu)$, is determined by a Maxwell construction (see Sec
\ref{sec_maxwell}). For
$N\rightarrow +\infty$, the density of states can be approximated
by
\begin{equation}
g(E)=e^{Ns[f_{1}]}+e^{Ns[f_{2}]}, \label{mb4b}
\end{equation}
where $f_{1}$ and $f_{2}$ are the distributions functions
corresponding to the global and local maxima of entropy (we shall
denote by $f_{*}$ the distribution function corresponding to the
saddle point of entropy). The corresponding temperatures
$\beta_{1}$, $\beta_{2}$ and $\beta_{*}$ form the series of
equilibria $\beta(E)$ in Fig. \ref{entropie}.  We now wish to
obtain the strict caloric curve $\beta_{micro}(E)$ defined in Sec.
\ref{sec_connexion}.
When $E$ is not too close from the transition energy $E_{t}$ and
$N\rightarrow +\infty$, we need just keep the contribution of the
global maximum of entropy. To investigate the situation close to the
transition energy, we rewrite the density of states (\ref{mb4b}) as
\begin{equation}
g(E)=e^{Ns[f_{1}]}\biggl\lbrack
1+e^{N(s[f_{2}]-s[f_{1}])}\biggr\rbrack, \label{mb4bi}
\end{equation}
Now, close to the transition point, we have
$s[f_1]=s[f_2]+\lambda'^{2}(\Lambda-\Lambda_{t})$, where $\lambda'$ is
a constant of order unity depending on $\mu$ (see
Fig. \ref{entropie}). Therefore,
\begin{equation}
g(E)=e^{Ns[f_{1}]}\biggl\lbrack
1+e^{-N\lambda^2\Delta\Lambda}\biggr\rbrack, \label{mb4bt}
\end{equation}
where $\Delta\Lambda=\Lambda-\Lambda_{t}$. Thus, the saddle point
approximation is valid for $N|\Lambda-\Lambda_{t}|\gg 1$. This
requires increasing large values of $N$ as we approach the
transition point $E_{t}$.

\section{The persistence of metastable states}
\label{sec_persistence}

\subsection{Typical lifetime of a metastable state}
\label{sec_metaeq}

In this section, we estimate the lifetime of a metastable state by
using an adaptation of the Kramers formula (Risken 1989).  We start
first by the canonical ensemble. Close to the critical temperature
$T_{c}$, see Fig. \ref{EbetaJMU1e3}, the free energy $F[f]$ has one
global minimum $F_{global}$ (stable GFEM), one local minimum
$F_{local}$ (metastable LFEM) and one saddle point $F_{saddle}$
(unstable SP). We call $E_{local}$ the energy of the metastable
equilibrium state and $E_{saddle}$ the energy of the unstable saddle
point. For a system initially prepared at $E_{local}$, the probability
of the energy $E$ is $P(E)\sim e^{-(F(E)-F_{local})/k_{B}T}$. Now, if
the energy fluctuations drive the system past $E_{saddle}$, it will
collapse. Therefore, the lifetime of the metastable state can be
estimated by $t_{life}\sim 1/P(E_{saddle})$ or
\begin{equation}
t_{life}\sim e^{\Delta F/k_{B}T},
\label{c1}
\end{equation}
where $\Delta F=|F_{saddle}-F_{local}|$ is the barrier of potential appropriate to our problem. Noting that
\begin{equation}
t_{life}\sim e^{N \Delta j},
\label{c2}
\end{equation}
we conclude that, except in the vicinity of the critical point
$T_{c}$, the lifetime of a metastable state scales as ${\rm
exp}(N)$. Therefore, metastable states (LFEM) are extremely robust in
astrophysics and cannot be neglected, even if there exists states with
lower free energy (GFEM). To investigate the situation close to the
critical point $T_{c}$, we shall calculate the barrier of potential
$\Delta j$ for the classical self-gravitating gas ($\hbar\rightarrow
0$). To that purpose, we use the results derived in a preceding paper
(Chavanis 2002a).

We recall that the series of equilibria is parameterized by
$\alpha=(4\pi G\beta\rho_{0})^{1/2}R$, where $\rho_{0}$ is the central
density. Introducing the Milne variables
\begin{equation}
u={\xi e^{-\psi}\over \psi'}, \qquad v=\xi\psi',
\label{milne}
\end{equation}
and noting $u_{0}=u(\alpha)$ and $v_{0}=v(\alpha)$ their value at the edge of the confining box, the thermodynamical parameters of the self-gravitating
gas are given by
\begin{equation}
\eta=v_{0},
\label{cc7}
\end{equation}
\begin{equation}
\Lambda={1\over v_{0}}\biggl ({3\over 2}-u_{0}\biggr ),
\label{cc8}
\end{equation}
\begin{equation}
s\equiv {S-S_{0}\over Nk_{B}}=-{1\over 2}\ln v_{0}-\ln (u_{0}v_{0})+v_{0}+2u_{0}-{3},
\label{cc9}
\end{equation}
where
\begin{equation}
{S_{0}\over Nk_{B}}=\ln\mu+{1\over 2}\ln\pi-\ln 2-{1\over 2}.
\label{cc9bis}
\end{equation}
The free energy per particle $j=s+\eta\Lambda$ is given by
\begin{equation}
j=-{1\over 2}\ln v_{0}-\ln(u_{0}v_{0})+v_{0}+u_{0}-{3\over 2}.
\label{c4}
\end{equation}

We now expand the Milne variables close to the critical point
$\alpha_{c}$. Recalling that $u(\alpha_{c})=1$,
$v(\alpha_{c})=\eta_{c}$ and $v'(\alpha_{c})=0$, we get
\begin{eqnarray}
u_{0}=1+{1\over\alpha_{c}}(2-\eta_{c})\epsilon-{1\over 2}{\eta_{c}\over\alpha_{c}^{2}}(2-\eta_{c})\epsilon^{2}\nonumber\\
+{1\over 6\alpha_{c}^{3}}(2-\eta_{c})(\eta_{c}^{2}+2\eta_{c}-4)\epsilon^{3}+...,
\label{c5}
\end{eqnarray}
\begin{eqnarray}
v_{0}=\eta_{c}+{\eta_{c}\over 2\alpha_{c}^{2}}(2-\eta_{c})\epsilon^{2}-{\eta_{c}\over 6\alpha_{c}^{3}}(2-\eta_{c})(2+\eta_{c})\epsilon^{3}+...,\nonumber\\
\label{c6}
\end{eqnarray}
where $\epsilon=\alpha-\alpha_{c}$. We also recall that
$\alpha_{c}\simeq 8.993195...$ and $\eta_{c}\simeq 2.517551...$. Substituting Eqs. (\ref{c5}) and (\ref{c6}) in  Eq. (\ref{c4}), we find that
\begin{eqnarray}
j=j_{0}-{1\over 4\alpha_{c}^{2}}(\eta_{c}-2)\epsilon^{2}+{1\over 12\alpha_{c}^{3}}(\eta_{c}-2)(10-3\eta_{c})\epsilon^{3}+...,\nonumber\\
\label{c7}
\end{eqnarray}
where $j_{0}=\eta_{c}-(3/2)\ln\eta_{c}-1/2$. On the other hand, recalling that $\eta=v(\alpha)$, we get
\begin{equation}
\eta_{c}-\eta={\eta_{c}(\eta_{c}-2)\over 2\alpha_{c}^{2}}\epsilon^{2}-{\eta_{c}(\eta_{c}^{2}-4)\over 6\alpha_{c}^{3}}\epsilon^{3}+...
\label{c8}
\end{equation}
Eliminating $\epsilon$ from the foregoing relations, we finally obtain
\begin{equation}
j=j_{0}+(\eta-\eta_{c})\biggl\lbrack K\pm\lambda(\eta_{c}-\eta)^{1/2}\biggr\rbrack
\label{c9}
\end{equation}
with $K=1/2\eta_{c}$ and
$\lambda=(2/3)\sqrt{2(\eta_{c}-2)/\eta_{c}^{3}}$. We note that
$K=\Lambda_{0}$ where $\Lambda_{0}$ is the normalized energy at the
critical temperature (this comes from the fact that $\delta
J=-E\delta\beta$). Equation (\ref{c9}) reproduces the cusp at
$\eta=\eta_{c}$ formed by the curve $j(\eta)$ in
Fig. \ref{EbetaJMU1e3}. Since $dJ=-Ed\beta$, $J(\alpha)$ and
$\eta(\alpha)$ are extrema at the same points in the series of
equilibria, which is at the origin of the cusp. From Eq. (\ref{c9}), the barrier of free energy near the critical point $\eta_{c}$ is given by
\begin{equation}
\Delta j=2\lambda(\eta_{c}-\eta)^{3/2}.
\label{c10}
\end{equation}
Therefore, the lifetime of the metastable state  scales as
\begin{equation}
t_{life}\sim e^{2\lambda N (\eta_{c}-\eta)^{3/2}}
\label{c11}
\end{equation}
with $\lambda\simeq 0.16979815...$. Therefore, metastable states are robust if
$N(\eta_{c}-\eta)^{3/2}\gg 1$. Note that if we had estimated the
lifetime of the metastable state by $t_{life}\sim e^{\Delta
E/k_{B}T}$, we would have obtained $t_{life}\sim e^{2\lambda'' N
(\eta_{c}-\eta)^{1/2}}$ with $\lambda''=\lbrack
2(\eta_{c}-2)/\eta_{c}\rbrack^{1/2}\simeq 0.64121317...$. We see that entropic
effects modify the power of $\eta_{c}-\eta$ in the expression of the
metastable state lifetime. Returning to Eq. (\ref{c11}), the
temperature of collapse $T_{l}$ taking into account finite $N$ effects can be
estimated by $2\lambda N (\eta_{c}-\eta_{l})^{3/2}\sim 1$. This leads to
\begin{equation}
\eta_{l}=\eta_{c}\biggl\lbrack 1-{1\over\eta_{c}}\biggl ({1\over 2\lambda}\biggr )^{2/3}N^{-2/3}\biggr\rbrack.
\label{c12}
\end{equation}
A numerical application gives
\begin{equation}
\eta_{l}=2.517\ (1-0.816\ N^{-2/3}).
\label{c12num}
\end{equation}

In the microcanonical ensemble, the lifetime of a metastable state can be estimated by
\begin{equation}
t_{life}\sim e^{\Delta S}\sim e^{N\Delta s},
\label{mc1}
\end{equation}
where $\Delta S$ is the entropic barrier.  By performing a study
similar to the previous one close to the turning point of energy
$\Lambda_{c}$ (see Fig. \ref{entropie}), we find that
\begin{equation}
s=s_{0}+(\Lambda_{c}-\Lambda)\biggl\lbrack \eta_{0}\pm \lambda'(\Lambda_{c}-\Lambda)^{1/2}\biggr\rbrack,
\label{mc2}
\end{equation}
with $s_{0}=-0.192962...$, $\eta_{0}=2.03085...$ and $\lambda'=0.863159...$. Therefore,
\begin{equation}
\Delta s=2\lambda'(\Lambda_{c}-\Lambda)^{3/2},
\label{mc3}
\end{equation}
and
\begin{equation}
t_{life}\sim e^{2\lambda'N(\Lambda_{c}-\Lambda)^{3/2}}.
\label{mc4}
\end{equation}
The energy of collapse $E_{l}$ taking into account finite $N$ effects can be estimated by $2\lambda'N(\Lambda_{c}-\Lambda)^{3/2}\sim 1$. This leads to
\begin{equation}
\Lambda_{l}=\Lambda_{c}\biggl\lbrack 1-{1\over\Lambda_{c}}\biggl ({1\over 2\lambda'}\biggr )^{2/3}N^{-2/3}\biggr\rbrack.
\label{mc5}
\end{equation}
A numerical application gives
\begin{equation}
\Lambda_{l}=0.335\ (1-2.077\ N^{-2/3}).
\label{mc6}
\end{equation}
This corresponds to a density contrast
\begin{equation}
{\cal R}_{l}=708.6\ (1-6.014\ N^{-1/3}).
\label{mc7}
\end{equation}
These results are similar to those found by Katz \& Okamoto (2000) by
analyzing the microcanonical fluctuations of isothermal spheres. In
particular, the scaling with $N$ is the same.  In the following
section, we apply their theory of fluctuations to the canonical
ensemble and show the relation with the preceding approach.

\subsection{Canonical fluctuations in isothermal spheres}
\label{sec_fluc}

The canonical partition function can be written
\begin{equation}
Z(\beta)=\int e^{J(E)}dE,
\label{f1}
\end{equation}
where $J(E)=S(E)-\beta E$. As before, we shall consider the situation where
$J(E)$ has two maxima (stable) and one minimum (unstable).  This corresponds to the caloric curve of Fig. \ref{EbetaJMU1e3}. We shall
be interested here by the metastable state, i.e. the {\it local}
maximum of $J(E)$. Thus, we eliminate ``by hands'' the contribution of
the global maximum in the integral. Expanding $J(E)$ around the local
maximum up to second order, we obtain
\begin{equation}
Z(\beta)=e^{J(E)}\int e^{{1\over 2}J''(E)(\delta E)^{2}}d(\delta E),
\label{f2}
\end{equation}
where $E$ now refers to the energy of the metastable state and $\delta E$ is a fluctuation around equilibrium. According to the foregoing equation,
the probability of the fluctuation $\delta E$ is given by
\begin{equation}
P(\delta E)\sim e^{{1\over 2}J''(E)(\delta E)^{2}}.
\label{f3}
\end{equation}
Noting that $J''(E)=S''(E)$ and recalling that ${\partial
S\over\partial E}=\beta(E)$ at equilibrium, we can rewrite the foregoing expression as
\begin{equation}
P(\delta E)={1\over\sqrt{2\pi}}\biggl |{d\beta\over dE}\biggr |^{1/2} e^{{1\over 2}{d\beta\over dE}(\delta E)^{2}}.
\label{f4}
\end{equation}
This formula shows that only equilibrium states with positive specific heats $C=dE/dT=-\beta^{2}{dE\over d\beta}>0$ are stable in the canonical ensemble. The variance of the fluctuations of energy is given by
the usual formula
\begin{equation}
\langle (\delta E)^{2}\rangle=-{1\over {d\beta\over dE}}=T^{2}C\ge 0,
\label{f5}
\end{equation}
which can also be directly derived from the exact $N$-body distribution function (\ref{can1a}). These results are valid on the whole gaseous  branch of
Fig. \ref{EbetaJMU1e3}. If we now examine the situation close to the
critical point ($\beta_{c}, E_{0}$) where $(d\beta/dE)_{c}=0$, we have
to first order
\begin{equation}
\biggl |{d\beta\over dE}\biggr |=\biggl |{d^{2}\beta\over dE^{2}}\biggr |_{c}(E-E_{0}).
\label{f6}
\end{equation}
We note $E'$ the energy of the unstable saddle point of free energy $J$ at temperature $T$. Close to the critical point ($\beta_{c}, E_{0}$) we can approximate the caloric curve $\beta(E)$ by a parabole. Thus, $E'$ is related to $E$,  the energy of the local maximum  of free energy $J$ at temperature $T$,  by
\begin{equation}
E-E'=2(E-E_{0}).
\label{f7}
\end{equation}
Using the criterion of Katz \& Okamoto (2000), adapted to the
canonical ensemble, we define the temperature of collapse as the
temperature $T_{l}$ at which the typical fluctuations of energy
$\delta E=\sqrt{\langle (\delta E)^{2}\rangle}$ are of the same order
as the difference $E-E'$. Indeed, as we approach the critical point
($\beta_{c}, E_{0}$) the fluctuations of energy become so important
(since the specific heat diverges) that the system can overcome the
barrier of potential played by the saddle point of free energy and collapse
(eventually reaching the global maximum of $J$). Thus, for finite $N$
systems, gravitational collapse can occur before the ending
($\beta_{c}, E_{0}$) of the metastable branch (spinodal
point). According to the preceding criterion, the temperature of
collapse is determined by the condition
\begin{equation}
\langle (\delta E)^{2}\rangle_{\beta_{l}}=4(E_{l}-E_{0})^{2}.
\label{f8}
\end{equation}
Using Eqs. (\ref{f5}) and (\ref{f6}), we obtain
\begin{equation}
E_{l}=E_{0}+ \biggl \lbrace 4\biggl |{d^{2}\beta\over dE^{2}}\biggr |_{c}\biggr \rbrace^{-1/3}.
\label{f9}
\end{equation}
It is more logical to write this criterion in terms of the temperature $T_{l}$. Close to the critical point, we have
\begin{equation}
\beta_{c}-\beta={1\over 2}\biggl |{d^{2}\beta\over dE^{2}}\biggr |_{c}(E-E_{0})^{2}.
\label{f10}
\end{equation}
Thus, using Eq. (\ref{f9}), we get
\begin{equation}
\beta_{l}=\beta_{c}-{1\over 8} \biggl \lbrace 4\biggl |{d^{2}\beta\over dE^{2}}\biggr |_{c}\biggr \rbrace^{1/3}.
\label{f11}
\end{equation}

\begin{figure}
\centering
\includegraphics[width=8.5cm]{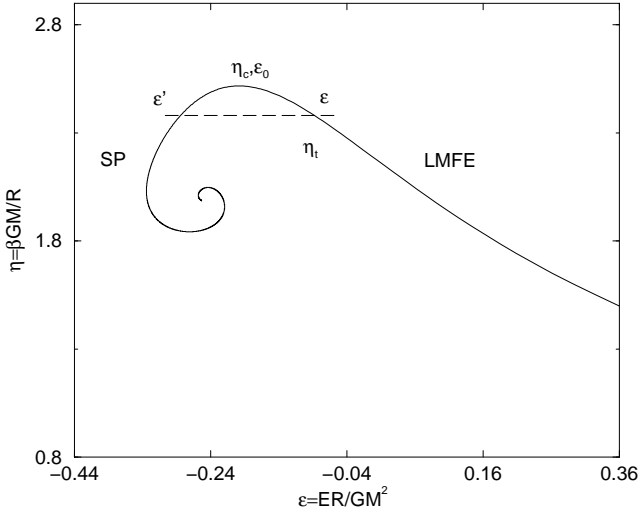}
\caption{Caloric curve showing metastable states (local
maxima of $J$) and unstable equilibria (saddle points of $J$) in
the canonical ensemble for classical particles ($\hbar\rightarrow
0$). As we approach the critical temperature $\eta_{c}$, the
fluctuations of energy increase considerably allowing the system
to collapse before the end of the metastable branch (spinodal
point). The temperature of collapse $T_{l}$ is the temperature
at which the typical fluctuations of energy are of the same order
as the energy difference $|\epsilon'-\epsilon|$.}
\label{collN}
\end{figure}

We can obtain more explicit results in the case of classical
isothermal spheres (see Fig. \ref{collN}). Introducing the
usual dimensionless parameters $\Lambda=-ER/GM^{2}$ and
$\eta=\beta GMm/R$, the foregoing relation can be rewritten
\begin{equation}
\eta_{l}=\eta_{c}-{1\over 8N^{2/3}} \biggl \lbrace 4\biggl |{d^{2}\eta\over d\Lambda^{2}}\biggr |_{c}\biggr \rbrace^{1/3},
\label{f12}
\end{equation}
Now, we have (see Chavanis 2002a)
\begin{equation}
\biggl ({d^{2}\eta\over d\Lambda^{2}}\biggr )_{c}=-{\eta_{c}^{3}\over\eta_{c}-2}.
\label{f13}
\end{equation}
Therefore, the temperature of collapse for finite $N$ systems is given by
\begin{equation}
\eta_{l}=\eta_{c}\biggl\lbrack 1-{1\over 8N^{2/3}}\biggl ({4\over \eta_{c}-2}\biggr )^{1/3}\biggr\rbrack.
\label{f14}
\end{equation}
A numerical application gives
\begin{equation}
\eta_{l}=2.517\ (1-0.247\ N^{-2/3}).
\label{f15}
\end{equation}
This estimate can be compared with Eq. (\ref{c12num}) which has the
same scaling with $N$. Of course, we should not give too much credit
on the numerical factor in front of $N^{-2/3}$ since the criterion for determining $T_{l}$  is
essentially phenomenological. The corresponding energy
$\Lambda_{l}$ can be deduced from Eq. (\ref{f10}) and is given by
\begin{equation}
\Lambda_{l}=\Lambda_{0}\biggl\lbrack 1- 2\biggl ({\eta_{c}-2\over 4}\biggr )^{1/3}N^{-1/3}\biggr \rbrack.
\label{f15b}
\end{equation}
It may also be of interest to determine the corresponding density
contrast ${\cal R}_{l}$. To that purpose, we start from the
formula (see Chavanis 2002a)
\begin{equation}
{\cal R}=e^{\psi(\alpha)}={\alpha^{2}\over u_{0}v_{0}},
\label{f16}
\end{equation}
and we use the expansion (\ref{c5}) and (\ref{c6}) of the Milne
variables. This yields
\begin{equation}
{\cal R}_{l}={\cal R}_{c}\biggl (1+{\eta_{c}\over\alpha_{c}}\epsilon_{l}\biggr ),
\label{f17}
\end{equation}
where ${\cal R}_{c}=\alpha_{c}^{2}/\eta_{c}$. Now, $\epsilon_{l}$ is determined by Eqs. (\ref{c8}) and (\ref{f14}) yielding
\begin{equation}
\epsilon_{l}={-\alpha_{c}\over 4^{1/3}N^{1/3}(\eta_{c}-2)^{2/3}}.
\label{f18}
\end{equation}
Substituting this result in Eq. (\ref{f17}) we finally obtain
\begin{equation}
{\cal R}_{l}={\cal R}_{c}\biggl \lbrace 1-{\eta_{c}\over \lbrack 2(\eta_{c}-2)\rbrack^{2/3}}N^{-1/3}\biggr\rbrace,
\label{f19}
\end{equation}
or, with numerical values,
\begin{equation}
{\cal R}_{l}=32.1\ (1-2.45\ N^{-1/3}).
\label{f20}
\end{equation}

The above theory thus predicts that, for finite $N$ systems, the
collapse should take place slightly before the canonical spinodal
point $\eta_{c}$ due to the enhancement of energy fluctuations as
we approach this critical point. The Monte Carlo simulations of de
Vega \& Sanchez (2002) in the canonical ensemble (with $N=2000$
particles) reveal that the collapse indeed takes place before the
critical point. However, the collapse occurs apparently at the
point where the isothermal compressibility
$\kappa_{T}=-(1/V)(\partial V/\partial p)_{T}$ diverges. This
corresponds to an inverse normalized temperature $\eta_{T}=2.43450...$.
It is sensibly smaller than the value obtained from Eq.
(\ref{f15}) with $N=2000$. The same discrepency with the
prediction of Katz \& Okamoto (2000) is found in the
microcanonical ensemble by de Vega \& Sanchez (2002). These
results seem to indicate that the higher collapse temperature and
energy found in Monte Carlo simulations are not due to finite $N$
effects. They seem to be independent on $N$ and correspond to
other critical points that do not coincide with the spinodal
point. We intend to perform independent Monte Carlo simulations to
check these results.

\subsection{General expression of the potential barrier}
\label{sec_app}

We can use the preceding approach to obtain a simple approximate
expression of the potential barrier close to the critical point
$\eta_c$. Consider a system at fixed temperature $T$ and
denote by $E$ its equilibrium energy such that $J'(E)=0$ (we
consider here that $E$ is the energy of the  metastable state).
For a fluctuation $E+\delta E$, we have seen that the variation of
free energy $\delta J={1\over 2}J''(E)(\delta E)^{2}$ can be
expressed as
\begin{equation}
\delta J={1\over 4}\beta''(E_{0})(E-E')(\delta E)^{2}
\label{ap1}
\end{equation}
where we have assumed that $E$ is close to the critical point $E_{0}$
and, as before, $E'$ is the energy of the minimum of $J$ at
temperature $T$.  We can use this expression to estimate the potential
barrier $\Delta J=J_{local}-J_{saddle}$. Thus, setting $\delta
E=E-E'$, we get
\begin{equation}
\Delta J={1\over 4}|\beta''(E_{0})|(E-E')^{3}.
\label{ap2}
\end{equation}
Using Eq. (\ref{f10}) to express this relation in terms of the
temperature, we finally obtain
\begin{equation}
\Delta J=\sqrt{32\over |\beta''(E_{0})|}(\beta_{c}-\beta)^{3/2} \qquad ({\rm approx.}).
\label{ap3}
\end{equation}
This is an estimate, because the curve $J(E)$ is not just a 
parabole between $E$ and $E'$.  Using Eq. (\ref{f13}), this
approximate expression can be written
\begin{equation}
\Delta j=6\lambda (\eta_{c}-\eta)^{3/2} \qquad ({\rm approx.}).
\label{ap4}
\end{equation}
It differs from the exact expression (\ref{c10}) by a factor $3$.
Noting that $\Delta J={1\over 2}(\Delta E)^{2}/\langle (\delta
E)^{2}\rangle$, according to Eqs.  (\ref{f4}) and (\ref{f5}), the
criterion (\ref{f8}) of Katz \& Okamoto (2000) corresponds to $\Delta
J\sim 1/2$. Alternatively, writing $t_{life}\sim e^{\Delta J}$, the
criterion leading to Eq. (\ref{c12}) corresponds to $\Delta J\sim 1$.
On a qualitative point of view, the approaches of
Secs. \ref{sec_metaeq} and \ref{sec_fluc} are equivalent. They
slightly differ on the details (definition of collapse temperature,
estimate of $\Delta J$) explaining why Eqs. (\ref{c12num}) and
(\ref{f15}) are not exactly the same.

We can also try to calculate $\Delta J$ by working directly on the
series of equilibria $\beta(E)$. Taking $E$ as a control parameter, we
have $J(E)=S(E)-\beta(E)E$. Noting that $J'(E)=-\beta'(E)E$,
$J''(E)=-\beta''(E)E-\beta'(E)$ and
$J'''(E)=-\beta'''(E)E-2\beta''(E)$, and expanding $J(E)$ close to the
critical point where $\beta'(E_{0})=0$, we get
\begin{eqnarray}
J(E)=J(E_{0})-{1\over
2}\beta''(E_{0})E_{0}(E-E_{0})^{2}\nonumber\\
-{1\over
3!}\lbrack \beta'''(E_{0})E_{0}+2\beta''(E_{0})\rbrack (E-E_{0})^{3}+... \label{ap5}
\end{eqnarray}
Using the relation
\begin{eqnarray}
\beta-\beta_{c}= {1\over 2}\beta''(E_{0})(E-E_{0})^{2}+{1\over 3!}\beta'''(E_{0})(E-E_{0})^{3}+...\nonumber\\
\label{ap6}
\end{eqnarray}
between the temperature and the energy close to the critical
point, we find that
\begin{eqnarray}
J(E)=J(E_{0})-E_{0}(\beta-\beta_{c})\nonumber\\
\pm {1\over 3} \beta''(E_{0})\biggl\lbrack {2(\beta-\beta_{c})\over \beta''(E_{0})}\biggr\rbrack^{3/2}+...
\label{ap6bis}
\end{eqnarray}
Therefore, close to the critical point, the barrier of free energy is
exactly given by
\begin{equation}
\Delta J={1\over 3}\sqrt{32\over |\beta''(E_{0})|}(\beta_{c}-\beta)^{3/2} \qquad ({\rm exact}),
\label{ap7}
\end{equation}
which returns Eq. (\ref{c10}). 

We can use the same type of approach in the microcanonical
ensemble  to obtain a simple approximate expression of the
entropic barrier close to the critical point $(\Lambda_c,\eta_0)$.
Consider a system at fixed energy $E$ and denote by $\beta$ its
equilibrium temperature (we
consider here that $\beta$ is the temperature of the metastable
state). For a fluctuation $\beta+\delta \beta$, the variation of
entropy can be expressed as (see Katz \& Okamoto 2000 for details)
\begin{equation}
\delta S={1\over 2}E'(\beta)(\delta\beta)^{2}.
\label{ap8}
\end{equation}
Assuming that $\beta$ is close to the critical point $\beta_{0}$,
and using arguments similar to those developed previously, we get
\begin{equation}
\delta S={1\over 4}E''(\beta_{0})(\beta-\beta')(\delta\beta)^{2}
\label{ap9}
\end{equation}
where $\beta'$ is the inverse temperature of the saddle point of
entropy.  We can use this expression to estimate the entropic barrier
$\Delta S=S_{local}-S_{saddle}$. Thus, setting $\delta
\beta=\beta-\beta'$, we get
\begin{equation}
\Delta S={1\over 4}E''(\beta_{0})(\beta-\beta')^{3}.
\label{ap10}
\end{equation}
Using
\begin{equation}
E-E_{c}\simeq {1\over 2}E''(\beta_{0})(\beta-\beta_{0})^{2},
\label{ap11}
\end{equation}
to express the temperature as a function of the energy close to
the critical point, we finally obtain
\begin{equation}
\Delta S=\sqrt{32\over E''(\beta_{0})}(E-E_{c})^{3/2} \qquad ({\rm approx.}).
\label{ap12}
\end{equation}
 We note that Eqs. (\ref{ap3}) and (\ref{ap12}) are symmetrical
 provided that we interchange $E$ and $\beta$. Evaluating numerically
 $d^{2}\Lambda/d\eta^{2}$ at the critical point
 $(\Lambda_{c},\eta_{0})$, this approximate expression can be written
\begin{equation}
\Delta s=6\lambda' (\Lambda-\Lambda_{c})^{3/2} \qquad ({\rm approx.}).
\label{ap13}
\end{equation}
It differs from the exact expression (\ref{mc3}) by a factor $3$. 

We can also try to calculate $\Delta S$ directly from the series of
equilibria $E(\beta)$. Taking $\beta$ as a control parameter, we have
$S(\beta)=J(\beta)+\beta E(\beta)$ and we recall that $dS=\beta dE$
and $dJ=-E d\beta$. Thus $S'(\beta)=\beta E'(\beta)$,
$S''(\beta)=E'(\beta)+\beta E''(\beta)$ and
$S'''(\beta)=2E''(\beta)+\beta E'''(\beta)$. Expanding $S(\beta)$
close to the critical point where $E'(\beta_{0})=0$, we get
\begin{eqnarray}
S(\beta)=S(\beta_{0})+{1\over
2}E''(\beta_{0})\beta_{0}(\beta-\beta_{0})^{2}\nonumber\\
+{1\over
3!}\lbrack 2E''(\beta_{0})+\beta_{0}E'''(\beta_{0})\rbrack (\beta-\beta_{0})^{3}+... \label{ap14}
\end{eqnarray}
Using the relation
\begin{eqnarray}
E-E_{c}= {1\over 2}E''(\beta_{0})(\beta-\beta_{0})^{2}+{1\over 3!}E'''(\beta_{0})(\beta-\beta_{0})^{3}+...\nonumber\\
\label{ap6gyd}
\end{eqnarray}
between the energy  and the temperature close to the critical
point, we find that
\begin{eqnarray}
S(\beta)=S(\beta_{0})+\beta_{0}(E-E_{c})\pm {1\over 3}E''(\beta_{0})\biggl\lbrack {2(E-E_{c})\over E''(\beta_{0})}\biggr\rbrack^{3/2}+...\nonumber\\
 \label{ap14bis}
\end{eqnarray}
Therefore, close to the critical point, the barrier of entropy is
exactly given by
\begin{equation}
\Delta S={1\over 3}\sqrt{32\over E''(\beta_{0})}(E-E_{c})^{3/2} \qquad ({\rm exact}),
\label{ap15}
\end{equation}
which returns Eq. (\ref{mc3}).

\section{Relation to the Kramers problem} \label{sec_kramers}

\subsection{The Fokker-Planck equation} \label{sec_fp}

In the preceding section, we have used the Kramers formula to estimate
the lifetime of metastable states in self-gravitating systems. We
would like now to {\it justify} this formula from first principles. In
order to determine the lifetime of a metastable state, we need to
introduce a dynamical model. In the canonical ensemble, we can
consider a system of self-gravitating Brownian particles (Chavanis,
Rosier \& Sire 2002) described by the stochastic equations
\begin{equation}
\label{fp1}
{d{\bf r}_{i}\over dt}={\bf v}_{i},
\end{equation}
\begin{equation}
\label{fp2}
{d{\bf v}_{i}\over dt}=-\nabla_{i}U({\bf r}_{1},...,{\bf r}_{N})-\xi{\bf v}_{i}+\sqrt{2D}{\bf R}_{i}(t),
\end{equation}
where $-\xi {\bf v}_{i}$ is a friction force and ${\bf R}_{i}(t)$ is a
white noise satisfying $\langle {\bf R}_{i}(t)\rangle={\bf 0}$ and
$\langle
{R}_{a,i}(t){R}_{b,j}(t')\rangle=\delta_{ij}\delta_{ab}\delta(t-t')$,
where $a,b=1,2,3$ refer to the coordinates of space and $i,j=1,...,N$
to the particles. The particles interact via the gravitational
potential $U({\bf r}_{1},...,{\bf r}_{N})=\sum_{i<j}u({\bf r}_{i}-{\bf
r}_{j})$ where $u({\bf r}_{i}-{\bf r}_{j})=-G/|{\bf r}_{i}-{\bf
r}_{j}|$. The inverse temperature $\beta=1/T$ is related to the
diffusion coefficient through the Einstein relation $\xi=D\beta$.

Using standard stochastic processes, we can derive the $N$-body
Fokker-Planck equation (Chavanis 2004b)
\begin{eqnarray}
\label{fp3}
{\partial P_{N}\over\partial t}+\sum_{i=1}^{N}\biggl ({\bf v}_{i}{\partial P_{N}\over\partial {\bf r}_{i}}+{\bf F}_{i}{\partial P_{N}\over\partial {\bf v}_{i}}\biggr )\nonumber\\
=\sum_{i=1}^{N} {\partial\over\partial {\bf v}_{i}}\biggl\lbrack D  {\partial P_{N}\over\partial {\bf v}_{i}}+\xi P_{N}{\bf v}_{i}\biggr\rbrack,
\end{eqnarray}
where ${\bf F}_{i}=-\nabla_{i}U({\bf r}_{1},...,{\bf r}_{N})$ is the gravitational force acting on particle $i$ and $P_{N}({\bf r}_{1},{\bf v}_{1},...,{\bf r}_{N},{\bf v}_{N},t)$ is the $N$-body distribution function. Its stationary states  correspond to the canonical distribution
\begin{eqnarray}
\label{fp4} P_{N}({\bf r}_{1},{\bf v}_{1},...,{\bf r}_{N},{\bf
v}_{N})={1\over Z(\beta)}e^{-\beta \bigl\lbrace
\sum_{i=1}^{N}{v_{i}^{2}\over 2}+U({\bf r}_{1},...,{\bf
r}_{N})\bigr\rbrace}.\nonumber\\
\end{eqnarray}
If we implement a mean-field approximation (Chavanis 2004b), we can
show that the distribution function $f({\bf r},{\bf v},t)=NP_{1}$ is
solution of the Kramers-Poisson system.  However, this is not the
approach that we shall consider here.

\begin{figure}
\centering
\includegraphics[width=8.5cm]{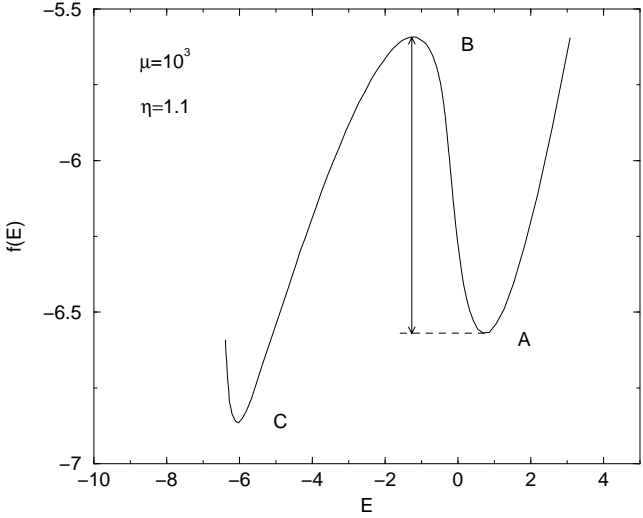}
\caption{Illustration of the barrier of free energy corresponding to
 Fig. \ref{ProbE}. Point $A$ corresponds to the gaseous
metastable state and point $B$ is the unstable solution creating a
barrier for reaching the condensed state $C$.  }
\label{EsMU10e3}
\end{figure}

We wish to obtain the time evolution of the distribution of energies
$P(E,t)$. To that purpose, we shall follow the method developed by
Kramers (1940) in his investigation of the escape of Brownian
particles over a potential barrier. The difference is that we work
here in a $6N$ dimensional phase space. Assuming that $P_{N}({\bf
r}_{1},{\bf v}_{1},...,{\bf r}_{N},{\bf v}_{N},t)$ depends only on
energy $E=\sum_{i=1}^{N}{v_{i}^{2}\over 2}+U({\bf r}_{1},...,{\bf
r}_{N})$ and time $t$, and averaging the Fokker-Planck equation
(\ref{fp3}) on the hypersurface of energy $E$, we show in Appendix B that
\begin{equation}
g(E){\partial P_{N}\over\partial t}(E,t)=3M{\partial\over\partial E}\biggl\lbrack I(E)\biggl ({\partial P_{N}\over\partial E}+\beta P_{N}\biggr )\biggr \rbrack,
\label{fp5}
\end{equation}
where $g(E)$ is the density of states and $I(E)$ is the phase space hypervolume with energy less than $E$ (thus $g(E)=dI/dE$). Now, the distribution of energies is given by
\begin{equation}
P(E,t)=P_{N}(E,t)g(E).
\label{fp6}
\end{equation}
At equilibrium, using Eq. (\ref{fp4}), we have
\begin{equation}
P(E)={1\over Z(\beta)}g(E)e^{-\beta E}.
\label{fp7}
\end{equation}
Out of equilibrium, substituting Eq. (\ref{fp6}) into Eq. (\ref{fp5}) and simplifying the resulting expressions, we finally obtain
\begin{equation}
{\partial P\over\partial t}={\partial\over\partial E}\biggl\lbrack D(E)\biggl ({\partial P\over\partial E}+\beta PF'(E)\biggr )\biggr \rbrack,
\label{fp8}
\end{equation}
where $D(E)=3M I(E)/g(E)$ and $F(E)=E-TS(E)=E-T\ln g(E)$ is the free
energy. This is similar to the Fokker-Planck equation describing the
stochastic motion of a particle in a potential where
the energy $E$ plays the role of the position $x$ and where the free
energy $F(E)$ plays the role of the potential $U(x)$. In the
following, we shall assume that the free energy $F(E)$ has a local
minimum at $E_{A}$ (metastable), a local maximum at $E_{B}$ (unstable)
and a global minimum at $E_{C}$. A typical situation is illustrated in
Fig. \ref{EsMU10e3}. We shall prepare a large number ${\cal N}\gg 1$
of systems close to the energy $E_{A}$ with the canonical distribution
(\ref{fp7}). Thus, ${\cal N}{\times} P(E,t)dE$ gives the number of
systems with energy between $E$ and $E+dE$ at time $t$.  As time goes
on, a fraction of these systems reaches the energy $E_{B}$ and
undergoes gravitational collapse towards $E_{C}$. Therefore, we adopt
the boundary condition
\begin{equation}
P(E_{B},t)=0.
\label{fp9}
\end{equation}
Our aim, now, is to estimate the current of diffusion past $E_{B}$ and the typical lifetime of metastable states.

\subsection{The stationary solutions} \label{sec_st}

The stationary solutions of Eq. (\ref{fp8}) are of the form
\begin{equation}
{\partial P\over\partial E}+\beta PF'(E)=-{J\over D(E)},
\label{st1}
\end{equation}
where $J<0$ is the current of diffusion in energy space. Using the boundary condition (\ref{fp9}), the solution of Eq. (\ref{st1}) reads
\begin{equation}
P(E)=J e^{-\beta F(E)}\int_{E}^{E_{B}}{e^{\beta F(x)}\over D(x)}dx.
\label{st2}
\end{equation}
The current of diffusion can therefore be expressed as
\begin{equation}
J={P_{A}e^{\beta F(E_{A})}\over\int_{E_{A}}^{E_{B}}{e^{\beta F(E)}\over D(E)}dE}.
\label{st3}
\end{equation}
To estimate the probability $P_{A}$, we shall approximate the curve $F(E)$ close to $A$ by a parabole. We thus make the expansion
\begin{equation}
F(E)=F(E_{A})+{1\over 2TC_{A}}(E-E_{A})^{2}+...
\label{st4}
\end{equation}
where we have used $F''(E)=-T\beta'(E)=1/TC$ where $C=dE/dT$ is the
specific heat. Therefore,
\begin{eqnarray}
P_{A}={1\over Z}e^{-\beta F(E_{A})}\qquad\qquad\qquad\qquad\qquad\nonumber\\
\simeq {1\over\int_{-\infty}^{+\infty}e^{-{\beta^{2}\over 2C_{A}}(E-E_{A})^{2}}dE}={\beta\over\sqrt{2\pi C_{A}}}.
\label{st5}
\end{eqnarray}
On the other hand, the integral in Eq. (\ref{st3}) is dominated by the value of the integrand close to $B$. Making the same quadratic expansion as in Eq. (\ref{st4}), we get
\begin{eqnarray}
\int_{E_{A}}^{E_{B}}{e^{\beta F(E)}\over D(E)}dE
\simeq -{1\over 2}{\times} {e^{\beta F(E_{B})}\over D(E_{B})}\int_{-\infty}^{+\infty}e^{{\beta^{2}\over 2C_{B}}(E-E_{B})^{2}}dE\nonumber\\
=-{e^{\beta F(E_{B})}\over 2 D(E_{B})}{\sqrt{2\pi |C_{B}|}\over\beta},\qquad
\label{st6}
\end{eqnarray}
where we recall that $C_{B}<0$ for the unstable solution. We thus obtain
the expression of the current
\begin{equation}
J=-{\beta^{2}D(E_{B})\over \pi \sqrt{C_{A}|C_{B}|}}  e^{-\beta (F_{B}-F_{A})}.
\label{st7}
\end{equation}
This expression involving the barrier of
free energy $\Delta F$, is similar to the one obtained by Kramers (1940) in his
classical study. In our case, the parameters have a thermodynamical
interpretation while Kramers considers a dynamical system in a
potential $U(x)$.

\subsection{The escape time} \label{sec_escape}

The preceding approach assumes that the population of systems that are
introduced at $A$ is continuously renewed so as to counterbalance the
population of systems that are lost at $B$ and maintain a stationary
regime. We shall now relax this simplifying assumption and look for
decaying solutions of Eq. (\ref{fp8}) of the form
\begin{equation}
P(E,t)=e^{-\lambda t}h(E),
\label{esc1}
\end{equation}
where  $h(E)$ satisfies the differential equation
\begin{equation}
{d\over dE}\biggl\lbrack D\biggl ({dh\over dE}+\beta h F'(E)\biggr )\biggr\rbrack=-\lambda h.
\label{esc2}
\end{equation}
Assuming that close to $E_{A}$ the system is at equilibrium, the foregoing equation can be integrated into
\begin{equation}
{dh\over dE}+\beta h F'(E)=-{\lambda \over D(E)}\int_{E_{A}}^{E} h(E')dE'.
\label{esc3}
\end{equation}
In usual situations, the eigenvalue $\lambda$ is expected to be small
as it corresponds to the inverse lifetime of the metastable states. We
thus consider the perturbative expansion of $h$ in powers of $\lambda$ and write
\begin{equation}
h=h_{0}+\lambda h_{1}+...
\label{esc4}
\end{equation}
Substituting this expansion  in Eq. (\ref{esc3}) and identifying terms of equal order, we obtain the differential equations
\begin{equation}
{dh_{0}\over dE}+\beta h_{0} F'(E)=0,
\label{esc5}
\end{equation}
\begin{equation}
{dh_{1}\over dE}+\beta h_{1} F'(E)=-{1 \over D(E)}\int_{E_{A}}^{E} h_{0}(E')dE'.
\label{esc6}
\end{equation}
The first equation integrates into
\begin{equation}
h_{0}=Ke^{-\beta F(E)}.
\label{esc7}
\end{equation}
Substituting this result in Eq. (\ref{esc6}), we obtain
\begin{equation}
{dh_{1}\over dE}+\beta h_{1} F'(E)=-{K \over D(E)}\int_{E_{A}}^{E} e^{-\beta F(E')} dE'.
\label{esc8}
\end{equation}
The solution of this first order differential equation can be written
\begin{equation}
h_{1}=-\chi(E)Ke^{-\beta F(E)},
\label{esc9}
\end{equation}
where the function $\chi$ is defined by
\begin{equation}
\chi'(E)={e^{\beta F(E)}\over D(E)}\int_{E_{A}}^{E}e^{-\beta F(E')}dE',
\label{esc10}
\end{equation}
with $\chi(E_{A})=0$. Therefore, in the approximation $\lambda\ll 1$, the solution of Eq. (\ref{fp8}) is
\begin{equation}
P(E,t)=Ke^{-\lambda t} e^{-\beta F(E)}\bigl\lbrack 1-\lambda \chi(E)\bigr\rbrack.
\label{esc11}
\end{equation}
The eigenvalue $\lambda$ is determined by the boundary condition (\ref{fp9}) yielding
\begin{equation}
\lambda={1\over\chi(E_{B})}.
\label{esc12}
\end{equation}
Therefore, the lifetime of the metastable state is given by
\begin{equation}
t_{life}\sim \chi(E_{B}).
\label{esc13}
\end{equation}
We can now try to simplify this expression. First, we approximate Eq. (\ref{esc10}) by
\begin{eqnarray}
\chi'(E)=-{e^{\beta F(E)}\over D(E)}\int_{-\infty}^{+\infty}e^{-\beta F(E_{A})}e^{{-\beta^{2}\over 2C_{A}}(E-E_{A})^{2}}dE\nonumber\\
=-{\sqrt{2\pi C_{A}}\over \beta}e^{-\beta F(E_{A})}{e^{\beta F(E)}\over D(E)}.\qquad
\label{esc14}
\end{eqnarray}
After integration, we obtain
\begin{equation}
\chi(E_{B})=-{\sqrt{2\pi C_{A}}\over \beta}e^{-\beta F(E_{A})}\int_{E_{A}}^{E_{B}}{e^{\beta F(E')}\over D(E')}dE'.
\label{esc15}
\end{equation}
With the additional approximation
\begin{eqnarray}
\chi(E_{B})={\sqrt{2\pi C_{A}}\over \beta}e^{-\beta F(E_{A})}\qquad\qquad\qquad\qquad\nonumber\\
{\times} {1\over 2}\int_{-\infty}^{+\infty}{e^{\beta F(E_{B})}\over D(E_{B})}e^{-{\beta^{2}\over 2|C_{B}|}(E-E_{B})^{2}}dE,
\label{esc17}
\end{eqnarray}
we finally get
\begin{eqnarray}
\lambda={\beta^{2}D(E_{B})\over \pi \sqrt{C_{A}|C_{B}|}}  e^{-\beta (F_{B}-F_{A})}.
\label{esc18}
\end{eqnarray}
Therefore, the lifetime of a metastable state behaves as
\begin{eqnarray}
t_{life}\sim e^{\Delta F\over kT}\sim e^{\Delta J}.
\label{esc18b}
\end{eqnarray}
We note that the expression of $\lambda$ is similar to the expression
(\ref{st7}) obtained for the current $J$. The connexion is the
following. In the non-stationary case, the current of diffusion at
$E_{B}$ is
\begin{eqnarray}
J_{B}=-D(E_{B})\biggl ({\partial P\over\partial E}+\beta PF'(E)\biggr )_{E_{B}}\nonumber\\
=-\lambda e^{-\lambda t} D(E_{B})\biggl ({\partial h_{1}\over\partial E}+\beta h_{1}F'(E)\biggr )_{E_{B}}\nonumber\\
=\lambda e^{-\lambda t} \int_{E_{A}}^{E_{B}} h_{0}(E')dE'\simeq
\lambda  \int_{E_{A}}^{E_{B}} P(E',t)dE'.
\label{esc19}
\end{eqnarray}
Hence, normalizing the current by the exponential decay of the density probability, we get
\begin{eqnarray}
{J_{B}\over e^{-\lambda t}}=-\lambda,
\label{esc20}
\end{eqnarray}
which is equivalent to Eq. (\ref{st7}).

In the preceding analysis, we have worked in the canonical ensemble
because the Brownian model (\ref{fp1})-(\ref{fp2}) is easier to study
than the $N$-stars Hamiltonian model, while exhibiting qualitatively
the same phenomena (phase transitions, metastable states etc.). We
expect to have symmetric expressions in the microcanonical ensemble
with the correspondance $E\leftrightarrow
\beta$ and $S\leftrightarrow J$. This study is left for a future work.

\section{Conclusion} \label{sec_conclusion}

In this paper, we have completed previous investigations concerning
the statistical mechanics of self-gravitating systems in
microcanonical and canonical ensembles. The microcanonical ensemble is
the proper description of isolated Hamiltonian systems such as
globular clusters (Binney \& Tremaine 1987). The canonical ensemble is
relevant for systems in contact with a heat bath of non-gravitational
origin. It is also the proper description of stochastically forced
systems such as self-gravitating Brownian particles (Chavanis, Rosier
\& Sire 2002).  We have justified the mean-field approximation, in a
proper thermodynamic limit $N\rightarrow +\infty$ with $\eta=\beta
GMm/R$ and $\epsilon=ER/GM^{2}$ fixed, from the equilibrium BBGKY
hierarchy. In this thermodynamic limit, the equilibrium state is
determined by a maximization problem: the maximization of entropy at
fixed mass and energy in the microcanonical ensemble and the
minimization of free energy at fixed mass and temperature in the
canonical ensemble. This determines the {\it most probable}
macroscopic distribution of particles at equilibrium. This can also be
seen as a saddle point approximation in the functional integral
formulation of the density of states and partition function. We have
shown that the saddle point approximation is less and less accurate
close to the transition point since the condition $N|E-E_{t}|\gg 1$
(in microcanonical ensemble) or $N|T-T_{t}|\gg 1$ (in canonical
ensemble) must be satisfied.

We have also argued that the lifetime of metastable states (local
entropy maxima) scales as ${\rm exp}(N)$ due to the long-range nature
of the interaction. Therefore, the importance of these metastable
states is considerable and they cannot be simply ignored. Metastable
states are in fact {\it stable} and they correspond to observed
structures in the universe such as globular clusters. The preceding
estimate must, however, be revised close to the critical point. By
solving a Fokker-Planck equation, we have shown the the lifetime of
metastable states is given by the Kramers formula involving the
barrier of entropy or free energy. These barriers have been calculated
exactly close to the Antonov energy $E_{c}$ (in microcanonical
ensemble) and close to the Jeans-Emden temperature $T_{c}$ (in
canonical ensemble). We have obtained the estimates $t_{life}\sim {\rm
exp}\lbrace 1.726\ N (\Lambda_{c}-\Lambda)^{3/2}\rbrace$ (in
microcanonical ensemble) and $t_{life}\sim {\rm exp}\lbrace 0.339\ N
(\eta_{c}-\eta)^{3/2}\rbrace$ (in canonical ensemble) so that the
lifetime decreases as we approach $E_c$ or $T_{c}$. This implies that
the collapse will take place slightly above $E_c$ or $T_{c}$ at an
energy $\Lambda_l=
\Lambda_{c}(1-2.077\ N^{-2/3})$ or temperature
$\eta_{l}=\eta_{c}(1-0.816\ N^{-2/3})$. Similar conclusions have been
reached by Katz \& Okamoto (2000). Yet, these predictions do not seem
to be consistent with the Monte Carlo simulations of de Vega \&
Sanchez (2002), although they find that the collapse indeed takes
place slightly before the critical point. Independent simulations are
under preparation to check that point.

Finally, a part of our discussion was devoted to answer the critics
raised by Gross (2003,2004) in recent comments. This author argues
that the microcanonical entropy $S_{micro}(E)$ and the microcanonical
temperature $\beta_{micro}(E)$ must be single valued. This is true in
a strict sense, but the problem is richer than that because of the
existence of long-lived metastable states. Therefore, the {\it
physical} caloric curve/series of equilibria $\beta(E)$ is
multi-valued and leads to ``dinosaur's necks'' and special
``microcanonical phase transitions'' (Chavanis 2002). This is specific
to systems with long-range interactions in view of the long lifetime
of metastable states (local entropy maxima). These results have
stimulated a general classification of phase transitions by Bouchet \&
Barr\'e (2004). Microcanonical phase transitions (as in
Fig. \ref{le5}) have not been fully appreciated by Gross and his
collaborators because their studies (e.g., Votyakov et al. 2002)
consider a {\it large} small-scale cut-off for which the caloric curve
looks like Fig. \ref{fel} and is univalued. If these authors reduce
their small-scale cut-offs, they will see ``dinosaurs'' appear!

\acknowledgements I am grateful to J. Katz for stimulating
discussions. While this paper was in course of redaction, I became
aware of a preprint by Antoni et al. [cond-mat/0401177] where similar
arguments about the lifetimes of metastable states for a toy model with
long-range interactions have been developed independently.

\appendix

\section{Justification of the mean-field approximation
from the equilibrium BBGKY hierarchy} 
\label{sec_bbgky}

In this Appendix, we show that the mean-field approximation is
exact for self-gravitating systems in a properly defined
thermodynamic limit $N\rightarrow +\infty$ with $\eta=\beta GMm/R$ and $\Lambda=-ER/GM^{2}$ fixed. In the canonical
ensemble, the equilibrium $N$-body distribution function is given by
\begin{equation}
P_{N}={1\over Z(\beta)}e^{-\beta U({\bf r}_{1},...,{\bf r}_{N})}, \label{w1}
\end{equation}
where we only consider the configurational part (the velocity part,
which is just a product of Maxwellians, is trivial). Here, $U({\bf
r}_{1},...,{\bf r}_{N})=\sum_{i<j}u_{ij}$ where $u_{ij}=-G/|{\bf
r}_{i}-{\bf r}_{j}|$ is the gravitational potential (we
can also use a soften potential in order to regularize the
partition function). Taking the derivative of Eq. (\ref{w1}) with
respect to ${\bf r}_1$, we get
\begin{equation}
{\partial P_{N}\over\partial {\bf r}_{1}}=-\beta P_{N}{\partial U\over\partial {\bf r}_{1}}. \label{w2}
\end{equation}
From this relation we can obtain the full equilibrium BBGKY hierarchy
for the reduced distribution functions (Chavanis 2004b).  Restricting
ourselves to the one and two-body distribution functions
\begin{equation}
P_{j}({\bf r}_{1},...,{\bf r}_{j})=\int P_{N}({\bf r}_{1},...,{\bf r}_{N})d^{3}{\bf r}_{j+1}...d^{3}{\bf
r}_{N}, \label{w2b}
\end{equation}
with $j=1,2$, we find
\begin{equation}
{\partial P_{1}\over\partial {\bf r}_{1}}({\bf r}_{1})=-\beta (N-1)\int P_{2}({\bf r}_{1},{\bf r}_{2}){\partial u_{12}\over\partial {\bf r}_{1}}d^{3}{\bf r}_{2}, \label{w3}
\end{equation}
\begin{eqnarray}
{\partial P_{2}\over\partial {\bf r}_{1}}({\bf r}_{1},{\bf r}_{2})=-\beta P_{2}({\bf r}_{1},{\bf r}_{2}){\partial u_{12}\over\partial {\bf r}_{1}}\qquad\qquad\qquad\nonumber\\
-\beta (N-2)\int P_{3}({\bf r}_{1},{\bf r}_{2},{\bf r}_{3}){\partial u_{13}\over\partial {\bf r}_{1}}d^{3}{\bf r}_{3}. \label{w4}
\end{eqnarray}
As is well-known, each equation of the hierarchy involves the next
order distribution function. We now decompose the two-body and
three-body distribution functions in the suggestive forms
\begin{equation}
P_{2}({\bf r}_{1},{\bf r}_{2})=P_{1}({\bf r}_{1})P_{1}({\bf r}_{2})+P_{2}'({\bf r}_{1},{\bf r}_{2}),\label{w5}
\end{equation}
\begin{eqnarray}
P_{3}({\bf r}_{1},{\bf r}_{2},{\bf r}_{3})=P_{1}({\bf r}_{1})P_{1}({\bf r}_{2})P_{1}({\bf r}_{3})+P_{2}'({\bf r}_{1},{\bf r}_{2})P_{1}({\bf r}_{3})\nonumber\\
+P_{2}'({\bf r}_{1},{\bf r}_{3})P_{1}({\bf r}_{2})+P_{2}'({\bf r}_{2},{\bf r}_{3})P_{1}({\bf r}_{1})+P_{3}'({\bf r}_{1},{\bf r}_{2},{\bf r}_{3}), \label{w6}
\end{eqnarray}
where $P_{n}'$ are the cumulants.  We shall consider the thermodynamic
limit $N\rightarrow +\infty$ with fixed $\eta=\beta GNm^{2}/R$. In
this limit, it can be shown that the non trivial correlations $P'_{n}$
are of order $N^{-(n-1)}$. Here, we shall just establish this result
for the two-body distribution function $P'_{2}$ .  Substituting the
decompositions (\ref{w5}) and (\ref{w6}) in Eqs. (\ref{w3}) and
(\ref{w4}) and assuming that $P'_{3}$ is negligible (this corresponds
to the Kirkwood approximation in plasma physics) the equation for the
two-body distribution function becomes after simplification
\begin{eqnarray}
{\partial P_{2}'\over\partial {\bf r}_{1}}({\bf r}_{1},{\bf r}_{2})=-\beta P_{1}({\bf r}_{1})P_{1}({\bf r}_{2}){\partial u_{12}\over\partial {\bf r}_{1}}\nonumber\\
-\beta P'_{2}({\bf r}_{1},{\bf r}_{2}){\partial u_{12}\over\partial {\bf r}_{1}}
-\beta N P'_{2}({\bf r}_{1},{\bf r}_{2})\int P_{1}({\bf r}_{3}){\partial u_{13}\over\partial {\bf r}_{1}}d^{3}{\bf r}_{3}\nonumber\\
-\beta NP_{1}({\bf r}_{1})\int P'_{2}({\bf r}_{2},{\bf r}_{3}){\partial u_{13}\over\partial {\bf r}_{1}}d^{3}{\bf r}_{3}. \label{w7}
\end{eqnarray}
We thus find that $P_{1}\sim 1$ and $P_2'\sim\beta u\sim \beta Gm^{2}/R= \eta/N=O(1/N)$. Therefore, in
the limit $N\rightarrow+\infty$, the two-body distribution
function is the product of two one-body distribution functions:
\begin{equation}
P_2({\bf r}_1,{\bf r}_2)=P_1 ({\bf r}_1)P_1 ({\bf r}_2)+O(1/N).
\label{w8}
\end{equation}
This justifies the exactness of the mean-field approximation for
self-gravitating systems. Note that $\eta/N$ can be identified as the
gravitational version of the ``plasma parameter''. This is similar to
the remark of Lundgren \& Pointin (1977) for the point vortex
gas. Now, plugging this result in Eq. (\ref{w3}), we find that, for
$N\rightarrow +\infty$,
\begin{equation}
{\partial P_{1}\over\partial {\bf r}_{1}}({\bf r}_{1})=-\beta NP_{1}({\bf r}_{1})\int P_{1}({\bf r}_{2}){\partial u_{12}\over\partial {\bf r}_{1}}d^{3}{\bf r}_{2}. \label{w9}
\end{equation}
Integrating with respect to ${\bf r}_1$ and introducing the mean density $\rho({\bf r})=\langle \sum_{i}m\delta({\bf r}_{i}-{\bf r})\rangle=NmP_{1}({\bf r})$, we obtain the Boltzmann
distribution
\begin{equation}
\rho=Ae^{-\beta m\Phi}, \label{w10}
\end{equation}
where $\Phi({\bf r})=\int \rho({\bf r}')u({\bf r}-{\bf r}')d^{3}{\bf
r}'$ is the self-consistent gravitational potential. Adding the gaussian velocity factor, we obtain the Maxwell-Boltzmann distribution
\begin{equation}
f=A'e^{-\beta m({v^2\over
2}+\Phi)}.\label{w11}
\end{equation}
As we have seen in Sec. \ref{sec_fermions}, the distribution function
(\ref{w11}) can also be obtained by minimizing the Boltzmann free
energy $F_B[f]$ at fixed mass $M$ and temperature $T$. This method
provides a condition of thermodynamical {stability} $\delta^{2}F\ge
0$, which is not captured by the equilibrium BBGKY hierarchy. To get
the condition of stability, we need to consider time-dependent
solutions, i.e. the non-equilibrium BBGKY hierarchy. Indeed, the
thermodynamical stability is related to the dynamical stability with
respect to the Fokker-Planck equation (Chavanis 2004c).

The equation for the two-body distribution function (\ref{w7}) is
complicated because the one-body distribution function is spatially
inhomogeneous. It may be of interest, however, to advocate the Jeans
swindle and consider, formally, the case of an infinite homogeneous
self-gravitating system (this can be made rigorous in a cosmological
context; see Kandrup 1983).  Making the drastic approximation $P_1=\rho/M$
where $\rho$ is a constant, Eq. (\ref{w7}) simplifies into
\begin{equation}
{\partial h\over\partial {\bf r}_{1}}({\bf r}_{1},{\bf r}_{2})=-\beta {\partial u_{12}\over\partial {\bf r}_{1}}-\beta {\rho\over m}\int h({\bf r}_{2},{\bf r}_{3}){\partial u_{13}\over\partial {\bf r}_{1}}d^{3}{\bf r}_{3}, \label{w12}
\end{equation}
where the second term in the right hand side of (\ref{w7}), of order
$1/N^{2}$, has been neglected. The correlation function $h$ is defined by
\begin{equation}
P_{2}(|{\bf x}|)={\rho^{2}\over M^{2}}\lbrack 1+h(|{\bf x}|)\rbrack, \label{w13}
\end{equation}
where ${\bf x}={\bf r}_{1}-{\bf r}_{2}$. Taking the divergence of
Eq. (\ref{w12}) and using $\Delta u=4\pi G m^{2}\delta({\bf r}_1-{\bf
r}_2)$, we obtain
\begin{equation}
\Delta h+k_{J}^{2} h=-4\pi G\beta m^{2}\delta({\bf x}), \label{w14}
\end{equation}
where $k_{J}=(4\pi G m\beta\rho )^{1/2}$ is the inverse of the Jeans
length. This equation is easily integrated to yield
\begin{equation}
h(x)=\beta G m^{2}{\cos(k_{J}x)\over x}.\label{w15}
\end{equation}
This is the counterpart of the Debye-H\"uckel result in the
gravitational case (Kandrup 1983). We emphasize that the above results
are valid for other systems with long-range interactions (Chavanis
2004b).  In particular, for the HMF model for which a homogeneous
phase rigorously exists, we find by the same method that
$h(\theta)={2\over N}{{\beta/\beta_{c}}\over
1-\beta/\beta_{c}}\cos\theta$, where $\beta_{c}={4\pi\over kM}$ is the
critical inverse temperature. In particular, the correlation function
diverges close to the critical point where the homogeneous phase
becomes unstable, so that the mean-field approximation ceases to be
valid. We expect a similar behavior for inhomogenous
self-gravitating systems close to $T_{c}$.

Considering now an isolated Hamiltonian system, the $N$-body
microcanonical distribution function is given by
\begin{eqnarray}
P_{N}({\bf r}_{1},{\bf v}_{1},...,{\bf r}_{N},{\bf v}_{N})={1\over g(E)}\delta (E-H({\bf r}_{1},{\bf v}_{1},...,{\bf r}_{N},{\bf v}_{N})).\nonumber\\
\label{w16}
\end{eqnarray}
From this expression it is easy to write the equilibrium BBGKY hierarchy (Chavanis 2004b). The first two equations of this hierarchy are 
\begin{eqnarray}
{\partial P_{1}\over\partial {\bf r}_{1}}(1)=-(N-1)\int {\partial u_{12}\over\partial {\bf r}_{1}} {1\over g(E)}{\partial \over\partial E}\biggl\lbrack g(E)P_{2}(1,2)\biggr\rbrack d^{3}(2),\nonumber\\
\label{w17}
\end{eqnarray}
\begin{eqnarray}
{\partial P_{2}\over\partial {\bf r}_{1}}({1},{2})=- {\partial u_{12}\over\partial {\bf r}_{1}}{1\over g(E)}{\partial\over\partial E}\biggl\lbrack g(E)P_{2}(1,2)\biggr\rbrack\qquad\qquad\qquad\nonumber\\
-(N-2)\int {\partial u_{13}\over\partial {\bf r}_{1}} {1\over g(E)}{\partial \over\partial E}\biggl\lbrack g(E)P_{3}(1,2,3)\biggr\rbrack d^{3}(3),\qquad \label{w18}
\end{eqnarray}
where we have written $(j)=({\bf r}_{j},{\bf v}_{j})$. Now,
\begin{equation}
{1\over g(E)}{\partial \over\partial E}\biggl\lbrack g(E)P_{j}\biggr\rbrack=\beta P_{j}+{\partial P_{j}\over\partial E}, \label{w19}
\end{equation}
where $\beta=\partial S/\partial E$ and $S(E)=\ln g(E)$. The ratio of
$\partial P_{j}/\partial E$ on $\beta P_{j}$ is of order ${1\over
E\beta}={1\over \Lambda\eta N}$. Therefore, in the thermodynamic limit
$N\rightarrow +\infty$ with $\Lambda$, $\eta$ fixed, the second term
in the r.h.s. of Eq. (\ref{w19}) is always negligible with respect to
the first. To leading order in $N$, we obtain the same
equations as in the canonical ensemble.
Therefore, the mean-field approximation is exact and leads to the
Boltzmann distribution (\ref{w10}). Observing that
\begin{eqnarray}
{\partial P_{1}\over\partial {\bf v}_{1}}(1)=- {1\over g(E)}{\partial \over\partial E}\bigl \lbrack g(E)P_{1}(1)\bigr \rbrack {\bf v}_{1},
\label{w20}
\end{eqnarray}
and taking the $N\rightarrow +\infty$ limit, we find that $P_{1}\sim
e^{\beta v_{1}^{2}/2}$. Combined with Eq. (\ref{w10}), this leads to
the Maxwell-Boltzmann distribution (\ref{w11}). Therefore, the
equilibrium BBGKY hierarchy in the microcanonical ensemble leads to
the same result (\ref{w11}) as in the canonical ensemble. As indicated
previously, the inequivalence of ensembles will appear by considering
the non-equilibrium BBGKY hierarchy. The thermodynamical stability in
the microcanonical ensemble is connected to the dynamical stability
with respect to the Landau equation (see Chavanis 2004c) which can be
deduced from the non-equilibrium BBGKY hierarchy to order $1/N$
(Balescu 1963).

\section{Derivation of Eq. (\ref{fp5})} \label{sec_der}

The phase space hypervolume with energy less than $E$ is defined by
\begin{eqnarray}
I(E)=\int_{H\le E}\prod_{i=1}^{N}d^{3}{\bf r}_{i}d^{3}{\bf v}_{i}.
\label{der1a}
\end{eqnarray}
Integrating over the velocities and using the fact that the kinetic term in the Hamiltonian is quadratic, a standard calculation yields
\begin{eqnarray}
I(E)=A\int \bigl \lbrack E-U({\bf r}_{1},...,{\bf r}_{N})\bigr \rbrack^{3N\over 2}\prod_{i=1}^{N}d^{3}{\bf r}_{i},
\label{der1}
\end{eqnarray}
where $A=(2/m)^{3N/2}V_{3N}$ and $V_{n}$ is the volume of a unit-hypersphere in a space of dimension $n$.
The density of states $g(E)=dI/dE$ is therefore given by
\begin{eqnarray}
g(E)={3N\over 2}A\int \bigl \lbrack E-U({\bf r}_{1},...,{\bf r}_{N})\bigr \rbrack^{{3N\over 2}-1}\prod_{i=1}^{N}d^{3}{\bf r}_{i}.
\label{der2}
\end{eqnarray}
Assuming now that $P_{N}({\bf r}_{1},{\bf v}_{1},...,{\bf r}_{N},{\bf
v}_{N},t)\simeq P_{N}(E,t)$, and substituting this ansatz in the
$N$-body Fokker-Planck equation Eq. (\ref{fp3}), we obtain after
simplification
\begin{eqnarray}
{\partial P_{N}\over\partial t}=2m\bigl\lbrack E-U({\bf r}_{1},...,{\bf r}_{N})\bigr \rbrack{\partial\over\partial E}\biggl (D{\partial P_{N}\over\partial E}+\xi P_{N}\biggr )\nonumber\\
+3Nm\biggl (D{\partial P_{N}\over\partial E}+\xi P_{N}\biggr ),\qquad\qquad\qquad
\label{der3}
\end{eqnarray}
where the term in bracket is $\sum_{i=1}^{N}v_{i}^{2}$. We note that
$P_{N}=P_{N}(E,t)$ is not an {\it exact} solution of (\ref{fp3}), as
expected. To get rid of the dependence in ${\bf r}_{1},...,{\bf
r}_{N}$, we shall average Eq. (\ref{der3}) over the hypersurface of
energy $E$ using
\begin{eqnarray}
\overline{X}(E)={\int (E-U)^{{3N\over 2}-1}X({\bf r}_{1},...,{\bf r}_{N}; E)\prod_{i=1}^{N}d^{3}{\bf r}_{i}\over \int (E-U)^{{3N\over 2}-1}\prod_{i=1}^{N}d^{3}{\bf r}_{i}},
\label{der4}
\end{eqnarray}
according to Eq. (\ref{der2}). This gives
\begin{eqnarray}
g(E){\partial P_{N}\over\partial t}=3M I(E){\partial\over\partial E}\biggl (D{\partial P_{N}\over\partial E}+\xi P_{N}\biggr )\nonumber\\
+3M g(E)\biggl (D{\partial P_{N}\over\partial E}+\xi P_{N}\biggr ).\qquad\qquad\qquad
\label{der5}
\end{eqnarray}
Using $g(E)=dI/dE(E)$, we can put this equation is the form (\ref{fp5}).

\section{Rotating self-gravitating systems} \label{sec_rot}

In this Appendix, we briefly consider the case of rotating
self-gravitating systems. Introducing dimensionless variables as
in Sec. \ref{sec_tl}, the conservation of angular momentum is
equivalent to ${\mb\lambda}\equiv {\bf
L}/(GM^{3}R)^{1/2}={\mb\lambda}[f']$ with
\begin{equation}
{\mb\lambda}[f']=\int {f'}{\bf r}'{\times}{\bf v}'  d^{3}{\bf
r}'d^{3}{\bf v}'. \label{rot1}
\end{equation}
Now, repeating the argumentation of Sec. \ref{sec_connexion}, the
density of states
\begin{eqnarray}
g(E,{\bf L})=\int \delta (E-H)\delta \biggl ({\bf L}-\sum_{i=1}^{N}m {\bf r}_{i}\times{\bf v}_{i}\biggr )\prod_{i=1}^{N}d^{3}{\bf r}_{i}d^{3}{\bf v}_{i},\nonumber\\
 \label{rot2}
\end{eqnarray}
can be written formally as
\begin{eqnarray}
g(E,{\bf L})=\int {\cal D}f \ e^{S[f]}
\delta(E-E[f])\nonumber\\
{\times}\delta(M-M[f])\delta({\bf L}-{\bf L}[f]). \label{rot3}
\end{eqnarray}
Similarly, the partition function
\begin{equation}
Z(\beta,{\bf \Omega})=\int e^{-\beta H+\beta{\bf\Omega}\cdot \sum_{i=1}^{N}m {\bf r}_{i}\times{\bf v}_{i}}\prod_{i=1}^{N}d^{3}{\bf r}_{i}d^{3}{\bf v}_{i} \label{rot4}
\end{equation}
can be written as
\begin{equation}
Z(\beta,{\bf \Omega})=\int e^{-\beta E+\beta{\bf\Omega}\cdot{\bf
L}}g(E,{\bf L})dEd^{3}{\bf L}, \label{rot5}
\end{equation}
or
\begin{equation}
Z(\beta,{\bf \Omega})=\int {\cal D}f \ e^{J[f]}\delta(M-M[f]),
\label{rot6}
\end{equation}
where $J[f]=S[f]-\beta E[f]+\beta {\mb\Omega}\cdot {\bf L}[f]$ is the
free energy.  In order to apply the saddle point approximation, we
just need to impose that $\Lambda=-ER/GM^2$, $\eta=\beta GMm/R$,
${\mb\lambda}={\bf L}/(GM^{3}R)^{1/2}$ and
${\bf\omega}={\bf\Omega}(R^{3}/GM)^{1/2}$ remain of order unity in the
limit $N\rightarrow +\infty$ (in the case of self-gravitating
fermions, we also need to impose that $\mu=\eta_0\sqrt{G^3 M R^3}$ is
fixed and in the case of a soften potential that $\epsilon=r_0/R$ is
fixed). This defines the thermodynamic limit for rotating
self-gravitating systems. The corresponding scalings are given in
Chavanis \& Rieutord (2003). In particular, $S\sim N$ and $J\sim
N$. Therefore, in the $N\rightarrow \infty$ limit, we have to maximize
$S[f]$ at fixed $E$, $M$ and ${\bf L}$ in the microcanonical ensemble
and we have to maximize $J[f]=S[f]-\beta E[f]+\beta\ {\bf\Omega}\cdot
{\bf L}[f]$ at fixed $\beta$, $M$ and ${\bf\Omega}$ in the canonical
ensemble. Computation of rotating self-gravitating systems in relation
with statistical mechanics have been performed by Votyakov et
al. (2002) for a classical gas on a lattice and by Chavanis \&
Rieutord (2003) for fermions.

\end{document}